\documentclass[12pt,twoside]{article}\usepackage{epsfig,rotate,colordvi}

\begin{document}
\newcommand{\np}{{\parindent0em\0}} 
\parskip3ex plus0.8ex minus0.8ex
\newcommand{\D}{displaystyle}
\newcommand{\T}{textstyle}
\newcommand{\TS}{scriptstyle}
\setlength{\topmargin}{-2cm}
\setlength{\textheight}{24cm}
\setlength{\headsep}{1.4cm}
\setlength{\voffset}{-0.6cm}

\pagestyle{plain}  

\newcommand{\ux}{\left}
\newcommand{\uy}{\right}

\newcommand{\ua}{\big}
\newcommand{\ub}{\Big}
\newcommand{\uc}{\bigg}
\newcommand{\ud}{\Bigg}

\newcommand{\vx}{\begin{array}{c}}
\newcommand{\vy}{\end{array}}

\newcommand{\w}{\frac}
\newcommand{\wx}{\begin{displaymath}}
\newcommand{\wy}{\end{displaymath}}

\newcommand{\x}{\large}
\newcommand{\xx}{\Large}
\newcommand{\xxx}{\LARGE}
\newcommand{\xxxx}{\huge}
\newcommand{\xxxxx}{\Huge}

\newcommand{\xy}{\normalsize}
\newcommand{\xyz}{\linebreak[4]}  
\newcommand{\zyx}{\newline}       

\newcommand{\xyx}{\underline}
\newcommand{\zyz}{\pagebreak[4]}

\newcommand{\yzy}{\zzzz{-0.4cm}\newpage} 

\newcommand{\yyz}[2]{\zzzz{#1cm}\zzzz{-0.4cm}\newpage\0 
\zzzz{#2cm}\zzzz{-1.3cm}} 


\newcommand{\y}{\small}
\newcommand{\yy}{\footnotesize}
\newcommand{\yyy}{\scriptsize}
\newcommand{\yyyy}{\tiny}

\newcommand{\z}{\mbox}
\newcommand{\zz}{\stackrel}
\newcommand{\zzz}{\hspace}
\newcommand{\zzzz}{\vspace}
\newcommand{\zzzzz}{\boldmath}
\newcommand{\zzzzzz}{\displaystyle} 

\renewcommand{\a}{\mbox{\"a}}   
\renewcommand{\o}{\mbox{\"o}}   
\renewcommand{\u}{\mbox{\"u}}   

\newcommand{\s}{\ss}  

\newcommand{\m}{\mbox{$\;\!\!$}}    
\newcommand{\mm}{\mbox{$\:\!\!$}}   
\newcommand{\mmm}{\mbox{$\!$}}      

\newcommand{\mmmm}{\mm\mm}          
\newcommand{\mmmmm}{\mmm\mm}        
\newcommand{\mmmmmm}{\mmm\mmm}      

\newcommand{\0}{\mbox{$\,\!$}}      

\newcommand{\1}{\mbox{$\:\!$}}      
\newcommand{\2}{\mbox{$\;\!$}}      
\newcommand{\3}{\mbox{$\,$}}        

\newcommand{\4}{\3\1}               
\newcommand{\5}{\3\2}               
\newcommand{\6}{\3\3}               

\newcommand{\7}{\4\3}               
\newcommand{\8}{\5\3}               
\newcommand{\9}{\6\3}               


\newcommand{\avec}[1]{\mbox{$#1_1,\ldots,#1_n$}}  
\newcommand{\avez}[1]{\mbox{$\overline{\zzz{#1cm}}$}} 

\newcommand{\bidir}[1]{ 
\z{\yyyy\zzzzz$         
\leftarrow\zzz{-0.288cm}\leftarrow\zzz{-0.288cm}\leftarrow\zzz{-0.288cm}
\leftarrow\zzz{-0.288cm}\leftarrow\zzz{-0.288cm}\leftarrow\zzz{-0.288cm}
\leftarrow\zzz{-0.288cm}\leftarrow\zzz{-0.288cm}\leftarrow\zzz{-0.288cm}
\leftarrow\zzz{-0.288cm}\leftarrow\zzz{-0.288cm}\leftarrow\zzz{-0.288cm}
\leftarrow\zzz{-0.288cm}\leftarrow\zzz{-0.288cm}\leftarrow\zzz{-0.288cm}
\leftarrow\zzz{-0.288cm}\leftarrow\zzz{-0.288cm}\leftarrow\zzz{-0.288cm}
\leftarrow
\zzz{-0.53cm}
\vx
\White{\z{\xxxxx\bf -}}\zzzz{-0.41cm}\\\White{\z{\xxxxx\bf -}}\zzzz{-0.41cm}\\
\White{\z{\xxxxx\bf -}}\zzzz{-0.08cm}
\vy
\zzz{-0.85cm}\zzz{#1cm}
\rightarrow\zzz{-0.288cm}\rightarrow\zzz{-0.288cm}\rightarrow\zzz{-0.288cm}
\rightarrow\zzz{-0.288cm}\rightarrow\zzz{-0.288cm}\rightarrow\zzz{-0.288cm}
\rightarrow\zzz{-0.288cm}\rightarrow\zzz{-0.288cm}\rightarrow\zzz{-0.288cm}
\rightarrow\zzz{-0.288cm}\rightarrow\zzz{-0.288cm}\rightarrow\zzz{-0.288cm}
\rightarrow\zzz{-0.288cm}\rightarrow\zzz{-0.288cm}\rightarrow\zzz{-0.288cm}
\rightarrow\zzz{-0.288cm}\rightarrow\zzz{-0.288cm}\rightarrow\zzz{-0.288cm}
\rightarrow
\zzz{-0.65cm}
\vx
\White{\z{\xxxxx\bf -}}\zzzz{-0.41cm}\\\White{\z{\xxxxx\bf -}}\zzzz{-0.41cm}\\
\White{\z{\xxxxx\bf -}}\zzzz{-0.08cm}
\vy
\zzz{-0.23cm}\zzz{-#1cm}\overline{\zzz{#1cm}} 
$}
} 

\newcommand{\mono}[1]{ 
\zzz{-2.5mm}
\unitlength0.05mm
\begin{picture}(#1,20)  
\put(0,10){\vector(1,0){#1}}
\end{picture}
\zzz{-2.5mm}
}

\newcommand{\duo}[1]{ 
\zzz{-2.5mm}
\unitlength0.05mm
\begin{picture}(#1,20)  
\put(0,10){\vector(1,0){#1}}
\put(#1,10){\vector(-1,0){#1}}
\end{picture}
\zzz{-2.5mm}
}

\newcommand{\poly}[2]   
{{
{{
\renewcommand{\drei}{\mono{#2}} 
\renewcommand{\vier}{\duo{#2}}  
\newcommand{\dreivier}[1]{##1}  
\dreivier{#1}                   
}}
}}

\newcommand{\drei}{
\unitlength0.05mm
\begin{picture}(20,33)
\put(-20,0){\z{\yyy\tt I\zzz{-0.07cm}I\zzz{-0.07cm}I }}
\put(-15.7,1){\line(1,0){50}}
\put(-15.7,2){\line(1,0){50}}
\put(-15.7,32){\line(1,0){50}}
\put(-15.7,33){\line(1,0){50}}
\end{picture}
}

\newcommand{\vier}{
\unitlength0.05mm
\begin{picture}(20,33)
\put(-20,0){\z{\yyy\tt \zzz{0.01cm}I\zzz{-0.04cm}V}}
\put(-15.7,1){\line(1,0){50}}
\put(-15.7,2){\line(1,0){50}}
\put(-15.7,32){\line(1,0){50}}
\put(-15.7,33){\line(1,0){50}}
\end{picture}
}

\newcommand{\eb}[2]{
\5\z{$\zzz{-0.28cm}\vx
\zzz{-0.2cm}\poly{#2}{66}\zzzz{-0.17cm}\\\z{#1}_{\zz{}{\5#2}}\zzzz{0.33cm}
\vy\zzz{-0.2cm}$}\5 
} 

\newcommand{\ebl}[2]{   
\5\z{$\zzz{-0.28cm}\vx
\zzz{-0.2cm}\poly{#2}{80}\zzzz{-0.1cm}\\\z{\x #1}_{\zz{}{\5#2}}\zzzz{0.4cm}
\vy\zzz{-0.2cm}$}\5 
} 

\newenvironment{p}[3]{\unitlength#1mm
\begin{picture}(#2,#3)}{\end{picture}}



\newcommand{\bp}{\begin{p}}  


\newcommand{\ep}{\end{p}}  

\newcommand{\pp}[4]{\put(#1,#2){\z{$\ux(\vx\zzz{#3cm}\zzzz{#4cm}\vy\uy)$}}}

\newcommand{\lp}[5]{\put(#1,#2){\line(#3,#4){#5}}}

\newcommand{\llp}{\thicklines}    

\newcommand{\und}{\z{\4,\4}}
\newcommand{\cit}[1]{\z{(\3#1\3)}} 

\newcommand{\defi}{\3:\zzz{0.02cm}=\3}   

\newcommand{\defin}{\z{$\zzzzzz
\zzz{0.02cm}\z{$\zzzzzz\vx\z{\yy\zzzzz$=$}\3\zzz{-0.36cm}
\White{\rule{0.08cm}{0.2cm}}\zzzz{0.02cm}\vy$}\zzz{-0.49cm}
\z{$\zzzzzz\vx\z{\yyy\zzzzz$:$}\zzzz{0.064cm}\vy\zzz{0.22cm}$}
$}}  

\newcommand{\op}[1]{     
\z{$\zzzzzz\zzz{-0.17cm}\vx
\epsfxsize#1cm\epsffile{shortGOTHICop.eps}
\zzzz{-0.15cm}\vy\zzz{-0.19cm}$}
}         

\newcommand{\ndouble}[1]{    
\z{ $\zzzzzz\zzz{-0.4cm}\vx
\epsfxsize#1cm\epsffile{angleN.eps}\zzzz{-0.04cm}\vy\zzz{-0.4cm}$ }
}      

\newcommand{\rdouble}[1]{    
\z{ $\zzzzzz\zzz{-0.4cm}\vx
\epsfxsize#1cm\epsffile{angleR.eps}\zzzz{-0.01cm}\vy\zzz{-0.4cm}$ }
}      

\newcommand{\anglea}[1] 
{\epsfxsize#1cm\epsffile{angleATTRIB.eps}}  

\newcommand{\minus}[2]{    
\z{$\zzzzzz
\z{\bf [}\6 
#1
\6\zzz{-0.18cm}\vx\z{\bf ,}\zzzz{-0.1cm}\vy\zzz{-0.18cm}\6
#2
\6{\z{\bf ]}}_{\zz{}{\z{\4\x\bf -}}}
$}
}

\newcommand{\plus}[2]{    
\z{$\zzzzzz
\z{\bf\{}\6 
#1
\8\zzz{-0.18cm}\vx\z{\bf ,}\zzzz{-0.1cm}\vy\zzz{-0.18cm}\8
#2
\6{\z{\bf\}}}
\zzz{-0.14cm}
\newcommand{\plu}{\z{$\zzzzzz
\renewcommand{\plu}{\z{$\zzzzzz 
\renewcommand{\plu}{\z{$\zzzzzz 
\zzz{-0.22cm}\z{\yyyy\bf +}$}}  
\z{\yyyy\bf +}\plu\plu\plu\plu\plu\zzz{-0.23cm}\z{\yyyy\bf +}$}}
\renewcommand{\plus}{\zzzz{-0.5cm}} 
\vx\plu\plus\\\plu\plus\\\plu\plus\\\plu\zzzz{-0.3cm}\vy$}}\plu
\zzz{-0.18cm}
$}
}

\newcommand{\unitv}[1]{  
\z{$\zzzzzz
{\z{\x\zzzzz$\hat{\zzz{-0.2mm}\z{\x\unboldmath$e$}\zzz{0.2mm}}$}}_
{\0\zz{}{\z{\yy#1}}}
$}                       
}

\newcommand{\katil}{ 
\z{$\zzzzzz
\zzz{-0.18cm}\vx\zzz{0.5mm}\z{\xx$\tilde{}$}\zzzz{-0.6cm}\\   
\z{\xy$\kappa$}\zzzz{-0.05cm}\vy\zzz{-0.18cm}                 
$}
}

\newcommand{\gast}{
\z{$\zzzzzz
\zzz{-0.18cm}\vx{\z{\y g}}^{\zz{\z{\y\zzzzz$\star$}}{}}  
\zzzz{-0.01cm}\vy\zzz{-0.18cm}
$}
}

\newcommand{\arrowone}[1]{\epsfysize#1mm\epsffile{arrow2.eps}}
\newcommand{\arrowtwo}[1]{\epsfysize#1mm\epsffile{arrow3a.eps}}
\newcommand{\arrowthree}[1]{\epsfysize#1mm\epsffile{arrow3b.eps}}

\newcommand{\uuuuu}{\unboldmath}

\newcommand{\n}{\zzz{0.05mm}}
\newcommand{\nn}{\zzz{0.1mm}}
\newcommand{\nnn}{\zzz{0.15mm}}
\newcommand{\nnnn}{\zzz{0.2mm}}
\newcommand{\nnnnn}{\zzz{0.25mm}}
\newcommand{\nnnnnn}{\zzz{0.3mm}}

\newcommand{\mn}{\zzz{-0.05mm}}
\newcommand{\mnn}{\zzz{-0.1mm}}
\newcommand{\mnnn}{\zzz{-0.15mm}}
\newcommand{\mnnnn}{\zzz{-0.2mm}}
\newcommand{\mnnnnn}{\zzz{-0.25mm}}
\newcommand{\mnnnnnn}{\zzz{-0.3mm}}

\newcommand{\vvx}{\zzz{-0.18cm}\vx}   
\newcommand{\vvy}{\vy\zzz{-0.18cm}}   
\newcommand{\vvv}[2]{   
\z{$\zzzzzz\vvx\z{\yyy$#1$}\zzzz{#2mm}\vvy$} }

\newcommand{\www}[2]{   
\z{$\zzzzzz\vvx\z{$#1$}\zzzz{#2mm}\vvy$} }

\newcommand{\uuu}[3]{{\renewcommand{\n}{\renewcommand}\renewcommand{\m}{\zzzzz}\n{\0}{\yyyy}\n{\1}{\yyy}\n{\2}{\yy}\n{\3}{\y}\n{\4}{\xy}\n{\5}{\x}\n{\6}{\xx}\n{\7}{\xxx}\n{\8}{\xxxx}\n{\9}{\xxxxx}\z{#1\n{\1}{\z{$\:\!$}}\n{\2}{\z{$\;\!$}}\n{\3}{\z{$\,$}}\n{\4}{\3\1}\n{\5}{\3\2}\n{\6}{\3\3}\n{\7}{\4\3}\n{\8}{\5\3}\n{\9}{\6\3}\n{\m}{\z{$\;\!\!$}}\renewcommand{\n}{\zzz{0.05mm}}$\zzzzzz\vvx\z{$#2$}\zzzz{#3mm}\vvy$}}}


\newcommand{\ra}{\z{$\rightarrow$}}  
\newcommand{\rra}{\z{$\longrightarrow$}}
\newcommand{\lra}{\z{$\leftrightarrow$}}
\newcommand{\llrra}{\z{$\longleftrightarrow$}}
\newcommand{\la}{\z{$\leftarrow$}}
\newcommand{\lla}{\z{$\longleftarrow$}}

\newcommand{\matrixm}{  
\z{\xy$\zzzzzz
\vvx M\zzzz{-0.19cm}\\\zzz{-0.1cm}
\z{\zzzzz$\widetilde{}$}\zzzz{-0.33cm}\vvy
$}
}

\newcommand{\sfdagger}[1]{ 
\z{\xy$\zzzzzz\vvv{\2
\z{\xy$\zzzzzz
\z{\zzzzz$|$}\zzz{-0.323cm}\vx\z{\xx -}\zzzz{0.1cm}\vvy 
$}
\2}{#1}$}
}   

\newcommand{\mroot}[2]{   
\z{\xy$\zzzzzz  
\z{\zzzzz$\zzzzzz\sqrt{\z{\uuuuu$\zzzzzz\zzz{-0.08cm}\zzz{#2mm}\zzz{#2mm}
\vx \zzzz{-0.48cm}\\#1\vy\zzz{-0.cm}\zzz{#2mm}\zzz{#2mm}$}}$}   
\zzz{-0.21cm}\vx\z{\yyyy\zzzzz$|\zzz{-0.098cm}|$}\zzzz{0.595cm}\zzzz{#2cm}
\vy\zzz{-0.21cm}\vx\z{\y\tt main}\zzzz{0.51cm}\zzzz{#2cm} \vvy
$}  
}   

\newcommand{\sffdagger}[1]{  
\z{\xy$\zzzzzz\vvx\2
\z{\xy$\zzzzzz
\z{\zzzzz$|$}\zzz{-0.323cm}\vx\z{\xx -}\zzzz{0.1cm}\vvy 
$}
\2#1\vvy$}
} 

\newcommand{\umu}[3]{   
\z{\xy$\zzzzzz 
\vvx
\Bigg(\zzzz{0.2cm}\vy\zzz{-0.17cm}\5
\z{\zzzzz$\zzzzzz\sqrt{\z{\uuuuu$\zzzzzz\zzz{-0.08cm}\zzz{#2mm}\zzz{#2mm}
\vx\zzzz{-0.48cm}\\{\z{U}}_{\zz{}{#1}}\zzzz{#3mm}\vy\zzz{-0.cm}\zzz{#2mm}
\zzz{#2mm}$}}$}\zzz{-0.21cm}\vx\z{\yyyy\zzzzz$|\zzz{-0.098cm}|$}
\zzzz{0.595cm}\zzzz{#2cm}\vy\zzz{-0.21cm}\vx\z{\y\tt main}\zzzz{0.51cm}
\zzzz{#2cm}\vy\zzz{-0.17cm}\6-\4\zzz{0.1mm}\vvx\ux[\zzz{0.2mm}
\zzz{-0.13cm}\vx\zzzz{-0.6cm}\\\z{\zzzzz$\zzzzzz\sqrt{\z{\uuuuu$\zzzzzz
\zzz{-0.08cm}\zzz{#2mm}\zzz{#2mm}\vx\zzzz{-0.48cm}\\{\z{U}}_{\zz{}{#1}}
\zzzz{#3mm}\vy\zzz{-0.cm}\zzz{#2mm}\zzz{#2mm}$}}$}\zzz{-0.21cm}\vx
\z{\yyyy\zzzzz$|\zzz{-0.098cm}|$}\zzzz{0.595cm}\zzzz{#2cm}\vy\zzz{-0.21cm}
\vx\z{\y\tt main}\zzzz{0.51cm}\zzzz{#2cm}\vy\zzz{-0.17cm}\zzzz{-0.2cm}
\vy\zzz{-0.11cm}\uy]\zzzz{0.1cm}\vy\zzz{-0.08cm}\mm
\sffdagger{\zzzz{7mm}\zzzz{#2cm}}\mnn\vx\Bigg)\zzzz{0.2cm}\vvy 
$}  
}   

\newcommand{\arra}[2]{\epsfxsize#1mm\epsfysize#2mm
\epsffile{arrow4.eps}}

\newcommand{\arrb}[2]{\epsfxsize#1mm\epsfysize#2mm
\epsffile{arrow5.eps}}

\newcommand{\arrc}[2]{\epsfxsize#1mm\epsfysize#2mm
\epsffile{arrow6.eps}}

\newcommand{\arrd}{   
\bp{1}{20}{10}\put(0,-2){\arra{20}{5.84}}\put(6.5,4.5){\arrc{7}{5}}\ep
}

\newcommand{\arre}{   
\bp{1}{10}{10}\put(0,-9){\arrb{5.84}{20}}\put(7,-2){\arrc{7}{5}}\ep
}

\newcommand{\auf}[1]{
\z{\epsfysize#1mm\epsffile{arrow7.eps}}
}   

\newcommand{\auff}[1]{
\z{\epsfysize#1mm\epsffile{arrow7a.eps}}
}   

\newcommand{\aufff}[1]{
\z{\epsfysize#1mm\epsffile{arrow7b.eps}}
}   


\newcommand{\drin}[1]{
\z{\epsfxsize#1mm\epsffile{arrow9.eps}}
}   


\newcommand{\zu}[1]{
\z{\epsfysize#1mm\epsffile{arrow8.eps}}
}   

\newcommand{\zuu}[1]{
\z{\epsfysize#1mm\epsffile{arrow8a.eps}}
}   

\newcommand{\zuuu}[1]{
\z{\epsfysize#1mm\epsffile{arrow8b.eps}}
}   


\newcommand{\intc}[1]{   
\z{\xy$\2\nnn\zzzzzz\www{\epsfxsize#1cm\epsffile{sum.eps}}{-0.5}\nnn\6$}
}


\newcommand{\intd}[1]{   
\z{\xy$\2\zzzzzz\www{\epsfxsize#1cm\epsffile{sum-integral.eps}}{-0.2}\6$}
}

\newcommand{\intA}[2]{   
\z{\xy$\4\n\zzzzzz
\www{\epsfxsize#1cm\epsfysize#2cm\epsffile{integral.eps}}{0.1}\nn\5$}
}
\newcommand{\intB}[2]{   
\z{\xy$\3\nnnn\zzzzzz
\www{\epsfxsize#1cm\epsfysize#2cm\epsffile{sum.upshape.eps}}{-0.5}\8$}
}
\newcommand{\intC}[2]{   
\z{\xy$\2\nnn\zzzzzz
\www{\epsfxsize#1cm\epsfysize#2cm\epsffile{sum.eps}}{-0.5}\nnn\6$}
}
\newcommand{\intD}[2]{   
\z{\xy$\2\zzzzzz
\www{\epsfxsize#1cm\epsfysize#2cm\epsffile{sum-integral.eps}}{-0.2}\6$}
}


\newcommand{\warning}[1]{   
\z{\bp{1}{#1}{#1}\put(1,-1){\epsfxsize#1mm\epsffile{warning.eps}}\ep\9}}

\newcommand{\calv}{\zzz{-0.1cm}\epsffile{calv.eps}} 

\newcommand{\implic}[2]{\epsfxsize#1mm\epsfysize#2mm\epsffile{implicat.eps}}

\newcommand{\implicat}[2]{\epsfxsize#1mm\epsfysize#2mm
\epsffile{implicat.bold.eps}}

\newcommand{\neque}{\z{$\xy   
\zzz{-0.3cm}\vx\zzz{0.2cm}\z{\yy\sf i.\5a.}\zzzz{-0.1cm}\\
{\x =}\z{\zzz{-0.22cm}\y\boldmath$\mid$}\zzzz{0.4cm}
\vy\zzz{-0.17cm}
$}}

\newcommand{\corr}{\z{$\5
\zzz{-0.47cm}\vx
\z{\xxxxx$\hat{\vx\zzzz{-1.66cm}\\\z{\xy$=$}\vy}$}
\zzzz{-0.32cm}\vy\zzz{-0.47cm}
\5$}}

\newcommand{\stroke}{\z{$
\Bigg|\9\zzz{-0.31cm}\Bigg|\9
$}}

\newcommand{\dyadc}{\z{$\3\zzz{-0.295cm}\vx\zzz{0.02cm}\epsfxsize0.4cm
\epsfysize0.2cm\epsffile{bishort.eps}\zzzz{-0.16cm}\\\z{C}\zzzz{0.324cm}
\vy\zzz{-0.22cm}$}}

\newcommand{\dyadx}{\z{$\3\zzz{-0.3cm}\vx\zzz{0.01cm}\epsfxsize0.36cm
\epsfysize0.12cm\epsffile{bishort.eps}\zzzz{-0.27cm}\\\z{x}\zzzz{0.232cm}
\vy\zzz{-0.23cm}$}}

\newcommand{\unit}[1]{
\zzz{-0.35cm}\zzz{-#1cm}\zzz{#1mm}\zzz{#1mm}\zzz{#1mm}\zzz{#1mm}
\zzz{#1mm}\zzz{#1mm}\zzz{#1mm}\zzz{#1mm}
\unitlength#1mm
\begin{picture}(60,60)
\thicklines

\multiput(0,0)(1,0){5}{
\multiput(0,28)(1,1){28}{\line(1,0){1}}
} 

\White{ 
\put(-48,0){\z{\yyyy\zzzzz$\zzzzzz 
\zzz{-0.25cm}\zzz{#1cm}\zzz{#1cm}\zzz{#1cm}\zzz{#1cm}\zzz{#1cm}
\vx
>\zzzz{-0.263cm}\\>\zzzz{-0.263cm}\\>\zzzz{-0.263cm}\\>\zzzz{-0.263cm}\\
>\zzzz{-0.263cm}\\>\zzzz{-0.263cm}\\>\zzzz{-0.263cm}\\>\zzzz{-0.263cm}\\
>\zzzz{-0.3cm}
\zzzz{#1cm}\zzzz{#1cm}\zzzz{#1cm}\zzzz{#1cm}\zzzz{#1cm}\zzzz{#1cm}
\vy$}}
}

\multiput(29,0)(1,0){2}{\line(0,1){57}}
\multiput(45,0)(1,0){2}{\line(0,1){57}}
\multiput(13,0)(0,1){2}{\line(1,0){50}}
\multiput(28,55)(0,1){2}{\line(1,0){18}}
 
\end{picture}
\zzz{-0.15cm}\zzz{#1mm}\zzz{#1mm}\zzz{#1mm}
}

\newcommand{\unitd}[1]{
\z{$\3\zzz{-0.295cm}\vx
\zzz{-#1cm}\epsfysize#1in\epsffile{bishort.eps}
\zzzz{-0.2cm}\zzzz{#1cm}\zzzz{-#1mm}\zzzz{-#1mm}\\
\zzz{0.15cm}\zzz{-#1cm}\unit{#1}\zzzz{0.09cm}\zzzz{#1in}\zzzz{#1in}
\vy\zzz{-0.23cm}\zzz{-#1in}\zzz{#1cm}\zzz{#1cm}$}
}

\newcommand{\dyada}{\z{$\3\zzz{-0.15cm}\vx\zzz{0.03cm}
\epsffile{bishort.eps}\zzzz{-0.26cm}\\\z{a}\zzzz{0.231cm}\vy\zzz{-0.11cm}$}}

\newcommand{\triada}{\z{$\zzz{-0.21cm}\vx\zzz{0.08cm}
\epsffile{trishort.eps}\zzzz{-0.35cm}\\\z{a}\zzzz{0.151cm}\vy\zzz{-0.115cm}$}}

%
%
\newcommand{\alphaS}{\alpha_s}
\newcommand{\alphaEM}{\alpha}
\newcommand{\PGE}{PGE}
\newcommand{\MSV}{SVM}
\newcommand{\pert}{P}
\newcommand{\nprt}{N\!P}
\newcommand{\SVM}{SVM}
\newcommand{\pbar}{{\bar{p}}}
\newcommand{\qbar}{{\bar{q}}}
\newcommand{\Jpsi}{J/\psi}
\newcommand{\VM}{{VM}}
\newcommand{\Reggeon}{I\!\!R}
\newcommand{\Pomeron}{I\!\!P}
\newcommand{\sigmaDP}{\sigma_{DP}}
\newcommand{\sigmaDD}{\sigma_{DD}}
%
\newcommand{\be}{\begin{equation}}
\newcommand{\ee}{\end{equation}}
\newcommand{\vw}[1]{\mbox{$\,#1\,$}}
\newcommand{\vww}[1]{\mbox{$\;#1\;$}}
\newcommand{\bea}{\begin{eqnarray}}
\newcommand{\eea}{\end{eqnarray}}
\newcommand{\benn}{\begin{displaymath}}
\newcommand{\eenn}{\end{displaymath}}
\newcommand{\beann}{\begin{eqnarray*}}
\newcommand{\eeann}{\end{eqnarray*}}
%
%
%
\newcommand{\barray}{\begin{array}}
\newcommand{\earray}{\end{array}}
\newcommand{\inv}{\frac{1}}
\newcommand{\di}[1]{${\renewcommand{\0}{\renewcommand}
\0{\2}{\z{2$\:\!$-dimensional}}
\0{\3}{\z{3$\:\!$-dimensional}}
\0{\4}{\z{4$\;\!$-dimensional}}
\0{\7}{\z{7-dimensional}}
\mbox{#1}}$}                      
\newcommand{\DI}[1]{\z{$D=#1$}}   
\newcommand{\DIM}[1]{\z{($\,D=#1\,$)}}  
\newcommand{\four}{4$\;\!$-}      
\newcommand{\gtsim}{\;\lower-0.45ex\hbox{$>$}\kern-0.77em\lower0.55ex\hbox{$\sim$}\;}
\newcommand{\ltsim}{\;\lower-0.45ex\hbox{$<$}\kern-0.77em\lower0.55ex\hbox{$\sim$}\;}
\newcommand{\fm}{\z{fm}}
\newcommand{\mb}{\z{mb}}
\newcommand{\nb}{\z{nb}}
\newcommand{\MeV}{\z{MeV}}
\newcommand{\GeV}{\z{GeV}}
\newcommand{\TeV}{\z{TeV}}
\newcommand{\G}{{\cal G}}       
\newcommand{\GG}{\hat{\cal{G}}} 
\newcommand{\Identity}{{1\!\rm l}}
\newcommand{\TransverseFourier}{{\cal F}_2}
\newcommand{\impactT}{{\cal T}}
\newcommand{\Pc}{{\cal P}}      
\newcommand{\Ps}{{\cal P}_S}    
\newcommand{\Tr}{\z{Tr}}    
\newcommand{\trace}{\z{\xy\sl trace}} 
\newcommand{\tr}{\z{tr}}    
\newcommand{\re}{\z{Re}}    
\newcommand{\im}{\z{Im}}    
\newcommand{\diff}{\z{d}}   
\newcommand{\diffA}{\z{d}\1\mathcal{A}} 
\newcommand{\vecA}{\1\vec{\mathcal{A}}} 
\newcommand{\diffV}{\z{d}\mathcal{V}}   
\newcommand{\q}[1]{\z{\x\ttfamily\itshape\m#1\4}} 
\newcommand{\A}{\q{\1A\mmm}} 
\newcommand{\F}{\q{\1F\mm}}  
\newcommand{\Order}{{\cal O}}   
\newcommand{\g}{\z{\m\ttfamily\itshape g\1}}
\newcommand{\Projector}{\z{P}} 
\newcommand{\tp}{\z{${\2}^{\z{\yyy t}}\0$}}
\newcommand{\partialslash}{\partial\!\!\!\!\!\!\not\,\,}
\newcommand{\pslash}{p\!\!\!\!\!\not\,\,}
\newcommand{\epsslash}{\epsilon\!\!\!\!\!\not\,\,}
\newcommand{\Dslash}{D\!\!\!\!\!\!\not\,\,}
\newcommand{\Gslash}{G\!\!\!\!\!\!\not\,\,}
\newcommand{\Qslash}{Q\!\!\!\!\!\!\not\,\,}
\newcommand{\GTslash}{G\!\!\!\!\!\!\!\!\!\not\,\,}
\newcommand{\kslash}{k\!\!\!\!\!\!\not\,\,}
\newcommand{\qslash}{q\!\!\!\!\!\not\,\,}
\newcommand{\rightslash}{\!\!\!\!\zz{\8\,\rightarrow}{\partialslash}\!\!}
\newcommand{\leftslash}{\!\!\!\!\zz{\8\,\leftarrow}{\partialslash}\!\!}
%

%
%
\newcommand{\befig}{\begin{figure}}
\newcommand{\efig}{\end{figure}}
\newcommand{\betab}{\begin{table}}
\newcommand{\etab}{\end{table}}
\newcommand{\dhline}{\hline\hline}

\newcommand{\fig}[3]{{#3{#1}{#2}}}
\newcommand{\ver}[2]{\epsfxsize#1mm\epsffile{Fig_5.#2.eps}}
\newcommand{\hor}[2]{\begin{rotate}[r]
{\epsfysize#1mm\epsffile{Fig_5.#2.eps}}\end{rotate}}

\newcommand{\fFig}[3]
{\epsfig{file=Fig_5.#2.eps,width=#1mm,angle=#3}}

\newcommand{\Fig}[3]
{\includegraphics[height=#1\textwidth,angle=#3]{Fig_5.#2.eps}}


\newcounter{saveno}
\newcounter{subequation}
\newcommand{\ar}[1]{\arabic{#1}}
\newcommand{\st}[1]{\stepcounter{#1}}
\newcommand{\stc}[2]{\setcounter{#1}{#2}}
\newcommand{\alphaon}{\renewcommand{\0}{\renewcommand}
\stc{saveno}{\ar{equation}}\stc{subequation}{0}\st{saveno}
\0{\theequation}{\ar{section}.\ar{saveno}\2\alph{subequation}}
\0{\be}{\st{subequation}\begin{equation}}
\0{\0}{\z{$\,\!$}}
}

\newcommand{\alphaoff}{\renewcommand{\0}{\renewcommand}
\stc{equation}{\ar{saveno}}
\0{\theequation}{\ar{section}.\ar{equation}}
\0{\be}{\begin{equation}}
\0{\0}{\z{$\,\!$}}
} 

\newcommand{\fs}{\zzz{0.16cm}} 

\renewcommand{\theequation}{\ar{section}.\ar{equation}}
\newcommand{\SEC}[1]{\section{#1}\stc{equation}{0}\stc{figure}{0}}
\newcommand{\sn}[1]{\z{Sec\n\_$\1$#1}}   
\newcommand{\SNN}[1]{\section*{#1}} 
\renewcommand{\thefigure}{\ar{section}.\ar{figure}}

%
%
%
%

\pagestyle{empty}
\bp{1}{1}{100}\put(-8,0){
\parbox{13.71cm}{
\begin{titlepage}
\y
\begin{flushright}
\0

\zzzz{0.5cm}
\begin{tabular}{l}
HD-THEP-03-51\\
\end{tabular}
\end{flushright}

\vspace*{0.9truecm} 

\begin{center}
\boldmath
{\Large\bf A Novel Ansatz for the Energy\1-Momentum 
Tensor on the Lattice\rule{0cm}{0.66cm}}\\ 
\vspace*{0.2truecm}
\unboldmath

\vspace*{1.cm} 

\smallskip
\begin{center}
{\sc 
{\large J.~Holk$^\diamond$
}}\\
\vspace*{2mm}
{\sl Institut f\"ur Theoretische Physik, Universit\"at Heidelberg\\
Philosophenweg 16\1, D\1-\169120 Heidelberg, Germany}
\end{center}

\vspace{2.0truecm}

{\large\bf Abstract\\[10pt]} \parbox[t]{\textwidth}{ 
\parbox{13.71cm}{
The comparison of structural analogies between the energy\1-\1momentum 
tensors in general relativity and in a gauge theory of Yang\1-Mills 
type is tentatively extended to lattice physics. These considerations 
are guiding to a new lattice model for the\xyz symmetric 
energy\1-\1momentum tensor \vw{{\Theta}_{\mm\zz{}{\mu\1\nu}}} 
of the pure Yang\1-Mills gauge sector,\xyz

\zzzz{-0.44cm}\np
basing on half powers of the plaquette variable. The concept of 
non\nnn-trivial princi\nnn-\xyz pal square roots of unitary matrices 
in lattice gauge theories can be epitomized to reconcile the 
pretension to a uniform construction principle for the components 
of \vw{{\Theta}_{\mm\zz{}{\mu\1\nu}}} with general qualitative 
thermodynamic demands concerning arguments in\xyz

\zzzz{-0.36cm}\np 
favour of a Wilson form 
for \vw{\zzzzzz 
\uuu{\5}{\langle}{0}\1\nnn
{\Theta}^{\zzz{0.01cm}\zzz{0.05mm}\zz{\z{\yyy$\mu$}}{}}
_{\zzz{-0.01cm}\zzz{0.05mm}\zz{\z{\yyy$\mu$}}{}}
\3\nnn\uuu{\5}{\rangle}{0}
} and \vw{\zzzzzz 
\langle\3
{\Theta}_{\mm\zz{}{4\14}}
\3\rangle
}. \z{\nn\0SU(2)\nn} 
Monte Carlo results for the\xyz

\zzzz{-0.4cm}\np
Euclidean expectation values 
on \z{a 
\z{$\4{\m1\mnn0}\mn\3\uuu{\4\m}{*\nnn*}{0.2}\mn\44\2\nnn$} 
lattice} 
are compared with that of com\nnn-\xyz peting hitherto existing lattice 
models for \vw{{\Theta}_{\mm\zz{}{\mu\1\nu}}}.
}}

\end{center}

\vskip 0.5 in

\zzzz{-0.5cm}
\footnoterule
\vskip 3truemm
\bp{0}{0}{1}\put(0,0){
{\small\tt 
\noindent$^\diamond$J.Holk@thphys.uni-heidelberg.de}
}\ep

\end{titlepage}
}
}\ep


\stc{page}{0}
\pagestyle{plain}
\newpage
 
%
%
%
%
%
%
%
%
%
%
\xy


\SEC{Introduction}
\label{Sec_Introduction}

An observable which comparatively little attention has so far been 
paid to is the energy\1-\1momentum tensor on the lattice. Apparently, 
a true deep un\nnn-\xyz derstanding of the energy\1-\1momentum tensor \vw{{\Theta}_{\mm\zz{}{\mu\1\nu}}} 
in general and its loose\xyz 

\zzzz{-1.06cm}\np
connection (\1the puzzle \z{\2'\1why\1?\1'\1} 
is still not yet resolved beyond comparisons regarding various 
applications of 
Noether's Theorem, cf\1.~\cite{GamboaSaravi:2003aq}\2) 
with general relativistic variational processes in continuum physics 
is still absent. If we cannot substantially improve this situation 
by standard classical, quantum, or lattice field theory we can instead 
look out for similar patterns in physics and try to explain something 
less understood by an unconventional confronta\1-\xyz tion with something 
else being less understood for similar reasons, thereby attempting to 
"\m\0cast out the devil by Beelzebub\2" in a positive sense. This is 
exactly what we will do in Section~2, inspiring a novel model for 
\vw{{\Theta}_{\mm\zz{}{\mu\1\nu}}} 
on the lattice.

With the help of several rather technical details and auxiliary 
definitions, to be given in Section~3, Section~4 will accurately 
specify various lattice mod\1-\xyz els for 
\vw{{\Theta}_{\mm\zz{}{\mu\1\nu}}}, 
including hitherto existing ones and the addressed new version.\xyz 

\zzzz{-1cm}\np
We will discuss results of Monte Carlo simulations in Section~5, 
regarding the Euclidean expectation values for the components of 
\vw{{\Theta}_{\mm\zz{}{\mu\1\nu}}} 
in the scope\xyz 

\zzzz{-1.02cm}\np
of the disposable lattice models. We will conclude with 
a final discussion in Section~6, where we will in particular focus 
upon general relativistic and thermodynamic aspects around 
\vw{{\Theta}_{\mm\zz{}{\mu\1\nu}}} 
and upon their 
implications for lattice\xyz

\zzzz{-1.01cm}\np
physics. Throughout the paper, bold type 
does not indicate vector rank.

\SEC{General Aspects}
\label{Sec_General_Aspects}

\zzzz{-0.2cm}
In Ref.~\cite{Holk:2003qa}\2, the energy\1-\1momentum tensors of 
general relativity, 
\vw{\Theta\vvv{\m\m\mu\1\nu}{-1.7}\zzz{-0.31cm}\vvv{(1)}{4.2}} 
(\1speci\nnn-\xyz

\zzzz{-1cm}\np  
fication by its equations of motion, being the Einstein field 
equations\1), and of\xyz

\zzzz{-1.04cm}\np  
a continuum Yang\1-Mills theory, 
\vw{\Theta\vvv{\m\m\mu\1\nu}{-1.7}\zzz{-0.31cm}\vvv{(2)}{4.2}} 
(\1representation by its own ingredients\1),\xyz

\zzzz{-1.01cm}\np
were shown to be 
analogously reducible to purely metric expressions and a\xyz

\zzzz{-1.02cm}\np
tensor field \vw{
\Omega\vvv{\4(k)}{5}\zzz{-0.46cm}
\vvv{\alpha\2\beta\2\n\gamma\2\nn\delta\1}{-3.2}
} for 
both \vww{
k\1\in\{\nnn1\2\mnn,2\1\n\}
} (\1see Eq.~(29) in Ref.~\cite{Holk:2003qa}\2\1--\1\vww{
\Theta\vvv{\m\m\mu\1\nu}{-1.7}\zzz{-0.31cm}\vvv{(2)}{4.2}}
there\xyz

\zzzz{-0.9cm}\np
corresponds to 
\vw{{\Theta}_{\mm\zz{}{\mu\1\nu}}} 
in \sn{1} here\1). It 
could be argued there that general\xyz

\zzzz{-1.02cm}\np
relativity automatically supplies 
an 
\z{\n\0SO\nn(4)\n} 
Yang\1-Mills tensor 
field \vw{
\Omega\vvv{\4(2)}{5}\zzz{-0.46cm}
\vvv{\alpha\2\beta\2\n\gamma\2\nn\delta\1}{-3.2}
} as\xyz

\zzzz{-1.01cm}\np
introduced just above, with

\zzzz{-0.66cm}
\be
\Omega\vvv{\4(2)}{5}\zzz{-0.46cm}
\vvv{\alpha\2\beta\2\n\gamma\2\nn\delta}{-3.2}
\8\4
\uuu{\2\m}{\propto}{1} 
\8\4
\Omega\vvv{\4(1)}{5}\zzz{-0.46cm}
\vvv{\alpha\2\beta\1\mu\2\n\nu}{-3.2}
\8
{g}^{\1\mu\1\rho}
\3
{g}^{\2\nu\1\sigma}
\8
\Omega\vvv{\4(1)}{5}\zzz{-0.46cm}
\vvv{\rho\2\sigma\2\gamma\2\nn\delta}{-3.2}
\ee

\zzzz{-0.55cm}\np
(v. Eq.~(36) in Ref.~\cite{Holk:2003qa}\2), revealing 
some kind of quadratic coherency 
between\xyz

\zzzz{-1.05cm}\np
\vw{
\Omega\vvv{\4(1)}{5}\zzz{-0.46cm}
\vvv{\alpha\2\beta\2\n\gamma\2\nn\delta\1}{-3.2}
} and \vw{
\Omega\vvv{\4(2)}{5}\zzz{-0.46cm}
\vvv{\alpha\2\beta\2\n\gamma\2\nn\delta\1}{-3.2}
}, both 
having the same symmetry properties 
(\1cf\1. Eq.~(30)\xyz

\yzy 

\zzzz{-0.88cm}\np
there\1).

Therefore it is possible to deduce from the 
vierbein calculus in Ref.~\cite{Holk:2003qa}\xyz

\zzzz{-1.04cm}\np
that \vw{
\Omega\vvv{\4(1)}{5}\zzz{-0.46cm}
\vvv{\alpha\2\beta\2\n\gamma\2\nn\delta\1}{-3.2}
} can be viewed in the context of referring to an 
\z{\n\0SO\nn(4)\n} field\xyz

\zzzz{-1.04cm}\np
strength tensor \vw{
{\z{\x F}}_{\zz{}{\zzz{-0.12cm}\z{\yyy\zzzzz$\mu\zzz{0.02cm}\nu$}}}
} while \vw{
\Omega\vvv{\4(2)}{5}\zzz{-0.46cm}
\vvv{\alpha\2\beta\2\n\gamma\2\nn\delta\1}{-3.2}
} (v. Eq.~(31), Ref.~\cite{Holk:2003qa}\2)  
can be related to\xyz

\zzzz{-0.91cm}\np
any Yang\1-Mills type field strength   
tensor\1; i.\1e. more roughly

\zzzz{-0.66cm}
\be
\Omega\vvv{\4(1)}{5}\zzz{-0.46cm}
\vvv{\alpha\2\beta\2\n\gamma\2\nn\delta}{-3.2}
\zzz{0.066cm}\6,\6
\Omega\vvv{\4(2)}{5}\zzz{-0.46cm}
\vvv{\alpha\2\beta\2\n\gamma\2\nn\delta}{-3.2}
\zzz{1cm}
\vx
\zzz{0.cm}\z{\tt continuum}\zzzz{-0.33cm}\\
\bp{1}{36}{1}
\llp
\put(18,0){\vector(-1,0){18}}
\put(18,0){\vector(1,0){18}}
\ep\zzzz{0.586cm}\\
\vy
\zzz{1cm}
{\z{\x F}}_{\zz{}{\zzz{-0.12cm}\z{\yyy\zzzzz$\mu\zzz{0.02cm}\nu$}}}
\z{\7.}
\ee

\zzzz{-0.68cm}\np
Regarding Formulae (31) up to (35) in Ref.~\cite{Holk:2003qa}\2, the 
allocation (2.2) becomes plausible relative to that context.

\zzzz{0.1cm}
In continuum physics, 
\vw{\Theta\vvv{\m\m\mu\1\nu}{-1.7}\zzz{-0.31cm}\vvv{(k)}{4.2}},
\vw{
\Omega\vvv{\4(k)}{5}\zzz{-0.46cm}
\vvv{\alpha\2\beta\2\n\gamma\2\nn\delta\1}{-3.2}
} (\1we 
are silently restricting ourselves\xyz

\zzzz{-0.95cm}\np
to \vww{
k\1\in\{\nnn1\2\mnn,2\1\n\}
} throughout the entire paper, as done in Ref.~\cite{Holk:2003qa}\2), 
and corre\2-\xyz

\zzzz{-0.96cm}\np
sponding Lagrangian densities \vw{
{\cal L}^{\zz{\z{\yy\1(\nn\z{\yyy\sf$k$})}}{}}
\mm} are of order 
\z{\1"\m$k$\2\nnnn"\m} 
in \vw{
{\z{\x F}}_{\zz{}{\zzz{-0.12cm}\z{\yyy\zzzzz$\mu\zzz{0.02cm}\nu$}}}
}. On a spacetime\xyz

\zzzz{-0.98cm}\np
lattice, however, the standard adaptations  
of \vw{
{\cal L}^{\zz{\z{\yy\1(\nn\z{\yyy\sf$k$})}}{}}
\mm} are 
known to be of order\xyz

\zzzz{-1.02cm}\np
one in the plaquette 
variable (\2\cite{Karsch:ve,Michael:1986yi,Rothe:1995hu} and 
v. Eq.~(2.25) in Ref.~\cite{Menotti:1986uc}\2), or\xyz in 
its respective substitute (\1being the 
so\nnn\1-\1called dual plaquette, e.\1g.--\nn\1cf\1. 
the\xyz

\zzzz{-1.02cm}\np
commentary on Eq.~(3.17) in Ref.~\cite{Smolin:1978kq}\2). In 
particular, the most transparent\xyz

\zzzz{-1.02cm}\np
lattice transference 
of \vw{
{\z{\x F}}_{\zz{}{\zzz{-0.12cm}\z{\yyy\zzzzz$\mu\zzz{0.02cm}\nu$}}}
} w.r.t. \vw{
\Omega\vvv{\4(1)}{5}\zzz{-0.46cm}
\vvv{\alpha\2\beta\2\n\gamma\2\nn\delta\1}{-3.2}
} in 
the framework of quantum gravity\xyz

\zzzz{-0.88cm}\np
(\mnn\1see Ref.~\cite{Smolin:1978kq}\2, 
focus on Formula (3.11) there and use it as a starting point\xyz

\zzzz{-1.04cm}\np
for introducing finite lattice spacings and 
constructing the dual plaquette if desired\1) suggests

\zzzz{-0.8cm}
\be
\z{R}\uuu{\1}{\1\z{a\1b}}{5}\zzz{-0.32cm}
\vvv{\mu\mnn\2\nu}{-2}
\2
(J\1,n\nn)
\8
\uuu{\2\m}{\propto}{1} 
\8
{
\Big(
\mmmmm
\begin{array}{l}
\zzz{0.4cm}\5
\z{\yy\1(\11\1)}
\zzzz{-0.26cm}\\
{\zz{\z{$\4\cal F\mmmm$}}{\z{\yy$\sim$}}}
\zz{}{\mmmmm\z{\yy\m$\mu\1\nu$}}
\zzzz{-0.cm}
\vy
\mmm
\m
\Big)
}
\mmmmm
\zz{\zz{\zz{\zz{\z{\yy a\1b}}{}}{}}{}}{}
\ee

\zzzz{-0.66cm}
\np 
with

\zzzz{-0.8cm}
\be
\1\mmm
\begin{array}{l}
\zzz{0.4cm}\5
\z{\yy \1(\11\1)}
\zzzz{-0.26cm}\\
{\zz{\z{$\4\cal F\mmmm$}}{\z{\yy$\sim$}}}
\zz{}{\mmmmm\z{\yy\m$\mu\1\nu$}}
\zzzz{-0.cm}
\vy\mmm\m
\2=\5
\vvx 
\z{\xx$\w{1}{\zzz{0.15mm}
\uuu{\3}{\12\4i\3g\m\mn\vvv{\circ}{-2.7}\2a^{\12}}{-1}}$}\zzzz{-0.9mm}
\zzzz{-0.3mm} 
\vy\vvx
\bigg(\zzzz{0.16cm}\vy\zzz{-0.17cm}
\5  
{\z{U}}_{\mm\zz{}{\z{\yy$\mu\1\nu$}}}
\5-\5
{\z{U}}_{\mm\zz{}{\z{\yy$\mu\1\nu$}}}\zzzz{-0.01cm}
\zzz{-0.11cm}  
\vx
\sfdagger{0}
\zzzz{0.5cm} 
\vy\zzz{-0.19cm} 
\0\vx
\bigg)\zzzz{0.16cm}\vvy 
\z{\7,}
\ee

\zzzz{-0.4cm}\np 
being equivalent to Formula (2.11) in Ref.~\cite{Menotti:1986uc} if 
\vw{\n\0g\zzz{-0.03cm}\vvv{\circ}{-2.7}} (\1there is of course no\xyz

\zzzz{-0.98cm}\np
relation with the metric tensor \vw{
g\m\vvv{\mu\1\nu\1}{-3}
}) is understood as the formally relevant\xyz

\zzzz{-0.96cm}\np
lattice coupling, 
\z{\1"\m$a$\2\n"\m} 
is the lattice spacing, 
\z{\1"\mnn$i$\3\mn"\m} 
is the imaginary unit, and\xyz

\zzzz{-1.02cm}\np
\vw{
{\z{U}}_{\mm\zz{}{\z{\yy$\mu\1\nu$}}}\n
} as usual denotes the plaquette variable.

\zzzz{0.2cm}
Now consider \vw{
\Omega\vvv{\4(2)}{5}\zzz{-0.46cm}
\vvv{\alpha\2\beta\2\n\gamma\2\nn\delta\1}{-3.2}
}. In principle there is no need to continue the 
pre\2-\xyz

\zzzz{-0.9cm}\np
cedingly 
prevailing assignment

\zzzz{-0.7cm}
\be
\Omega\vvv{\4(1)}{5}\zzz{-0.46cm}
\vvv{\alpha\2\beta\2\n\gamma\2\nn\delta}{-3.2}
\zzz{1cm}
\vx
\zzz{0.cm}\z{\tt lattice}\zzzz{-0.33cm}\\
\bp{1}{36}{1}
\llp
\put(18,0){\vector(-1,0){18}}
\put(18,0){\vector(1,0){18}}
\ep\zzzz{0.586cm}\\
\vy
\zzz{1cm}
\zzz{-0.18cm}
\begin{array}{l}
\zzz{0.4cm}\5
\z{\yy \1(\11\1)}
\zzzz{-0.26cm}\\
{\zz{\z{$\4\cal F\mmmm$}}{\z{\yy$\sim$}}}
\zz{}{\mmmmm\z{\yy\m$\mu\1\nu$}}
\zzzz{-0.cm}
\vy\zzz{-0.18cm}
\ee

\zzzz{-0.68cm}\np
to this type of separate situation. If we consistently define the 
main (--\2or\xyz

\zzzz{-1.02cm}\np
principal\1--) root \vw{
\mroot{M\1}{0}
} of any diagonalizable matrix \vw{M} (\nnn\0including the \xyz

\yzy 

\zzzz{-0.98cm}\np
special case of scalars given by 
\z{$1\2\uuu{\1}{\z{\sf X}}{0.5}\21$} 
matrices) to be performed by effectively\xyz

\zzzz{-1.02cm}\np
replacing its eigenvalues 
by the principal values of their own square 
roots\xyz \z{(\1v\mm.\1i\nn. in \sn{3}\2)} then 
we can introduce the competing lattice expression

\zzzz{-0.77cm}
\be
\1\mmm
\begin{array}{l}
\zzz{0.4cm}\5
\z{\yy \1(\12\1)}
\zzzz{-0.26cm}\\
{\zz{\z{$\4\cal F\mmmm$}}{\z{\yy$\sim$}}}
\zz{}{\mmmmm\z{\yy\m$\mu\1\nu$}}
\zzzz{-0.cm}
\vy\mmm\m
\1\defin\4
\vvx
\z{\xx$\w{1}{
\uuu{\3}{\1i\3g\m\mn\vvv{\circ}{-2.7}\2a^{\12}}{-1}}$}\zzzz{-0.9mm}
\zzzz{-0.3mm} 
\vy\vvx
\Bigg(\zzzz{0.2cm}\vy\zzz{-0.17cm}
\5
\z{\zzzzz$\zzzzzz\sqrt{\z{\unboldmath$\zzzzzz
\zzz{-0.08cm}\vx
\zzzz{-0.48cm}\\
{\z{U}}_{\mm\zz{}{\z{\yy$\mu\1\nu$}}}\zzzz{-0.01cm}
\vy\zzz{-0.cm}
$}}$}
\zzz{-0.21cm}\vx
\z{\yyyy\zzzzz$|\zzz{-0.098cm}|$}\zzzz{0.595cm}    
\vy\zzz{-0.21cm}\vx
\z{\y\tt main}\zzzz{0.51cm}  
\vy\zzz{-0.17cm}
\6
-
\4\zzz{0.1mm}
\vvx
\nn\m 
\zzz{0.2mm}
\zzz{-0.13cm}\vx
\zzzz{-0.6cm}\\
\z{\zzzzz$\zzzzzz\sqrt{\z{\uuuuu$\zzzzzz
\zzz{-0.08cm}\vx
\zzzz{-0.48cm}\\
{\z{U}}_{\mm\zz{}{\z{\yy$\mu\1\nu$}}}\zzzz{-0.01cm}
\vy\zzz{-0.cm}
$}}$}
\zzz{-0.21cm}\vx
\z{\yyyy\zzzzz$|\zzz{-0.098cm}|$}\zzzz{0.595cm}    
\vy\zzz{-0.21cm}\vx
\z{\y\tt main}\zzzz{0.51cm}  
\vy\zzz{-0.17cm}
\zzzz{-0.2cm}
\vy\zzz{-0.11cm}
\m 
\zzzz{0.1cm}
\vvy
\zzz{-0.08cm}\vx
\sfdagger{0}
\m 
\zzzz{0.7cm}
\vy\zzz{-0.19cm}
\2\vx
\Bigg)\zzzz{0.2cm}\vvy
\ee

\zzzz{-0.59cm}
\np
as well, 
with \vww{\zzzzzz
\lim_{a\1\rightarrow\10}\5
\zzz{-0.2cm}
\begin{array}{l}
\zzz{0.4cm}\5
\z{\yy\1(\nnn\1$k$\2\mn)}
\zzzz{-0.26cm}\\
{\zz{\z{$\4\cal F\mmmm$}}{\z{\yy$\sim$}}}
\zz{}{\mmmmm\z{\yy\m$\mu\1\nu$}}
\zzzz{-0.cm}
\vvy
\1\nnn=\2
{\z{\x F}}_{\zz{}{\zzz{-0.12cm}\z{\yyy\zzzzz$\mu\zzz{0.02cm}\nu$}}}
\zzz{0.4cm}
\forall\6\4\6k\9\in\6
\{\2\z{$1\2,\12\2\}\1$}
} and\zyx

\zzzz{-0.8cm}
\np
\z{\0\zzz{-0.3cm}
\vww{\zzzzzz
\vvx
\nn\m 
\zzz{0.2mm}
\zzz{-0.13cm}\vx
\zzzz{-0.6cm}\\
\z{\zzzzz$\zzzzzz\sqrt{\z{\uuuuu$\zzzzzz
\zzz{-0.08cm}\vx
\zzzz{-0.48cm}\\
{\z{U}}_{\mm\zz{}{\z{\yy$\mu\1\nu$}}}\zzzz{-0.01cm}
\vy\zzz{-0.cm}
$}}$}
\zzz{-0.21cm}\vx
\z{\yyyy\zzzzz$|\zzz{-0.098cm}|$}\zzzz{0.595cm}    
\vy\zzz{-0.21cm}\vx
\z{\y\tt main}\zzzz{0.51cm}  
\vy\zzz{-0.17cm}
\zzzz{-0.2cm}
\vy\zzz{-0.11cm}
\m 
\zzzz{0.1cm}
\vvy
\zzz{-0.08cm}\vx
\sfdagger{0}
\m 
\zzzz{0.7cm}
\vvy
\3
\defin 
\3
\www{\uc(}{2}
\2\nnn 
\vvx
\nn\m 
\zzz{0.2mm}
\zzz{-0.13cm}\vx
\zzzz{-0.6cm}\\
\z{\zzzzz$\zzzzzz\sqrt{\z{\uuuuu$\zzzzzz
\zzz{-0.08cm}\vx
\zzzz{-0.48cm}\\
{\z{U}}_{\mm\zz{}{\z{\yy$\mu\1\nu$}}}\zzzz{-0.01cm}
\vy\zzz{-0.cm}
$}}$}
\zzz{-0.21cm}\vx
\z{\yyyy\zzzzz$|\zzz{-0.098cm}|$}\zzzz{0.595cm}    
\vy\zzz{-0.21cm}\vx
\z{\y\tt main}\zzzz{0.51cm}  
\vy\zzz{-0.17cm}
\zzzz{-0.2cm}
\vy\zzz{-0.11cm}
\m 
\zzzz{0.1cm}
\vvy
\mm\vx
\uc)\zzzz{0.2cm}\vvy
\zzz{0.04cm}
\sfdagger{9.5}
}.}

\zzzz{0.3cm}
Demanding

\zzzz{-0.73cm}
\be
\Omega\vvv{\4(2)}{5}\zzz{-0.46cm}
\vvv{\alpha\2\beta\2\n\gamma\2\nn\delta}{-3.2}
\zzz{1cm}
\vx
\zzz{0.cm}\z{\tt lattice}\zzzz{-0.33cm}\\
\bp{1}{36}{1}
\llp
\put(18,0){\vector(-1,0){18}}
\put(18,0){\vector(1,0){18}}
\ep\zzzz{0.586cm}\\
\vy
\zzz{1cm}
\zzz{-0.18cm}
\begin{array}{l}
\zzz{0.4cm}\5
\z{\yy \1(\12\1)}
\zzzz{-0.26cm}\\
{\zz{\z{$\4\cal F\mmmm$}}{\z{\yy$\sim$}}}
\zz{}{\mmmmm\z{\yy\m$\mu\1\nu$}}
\zzzz{-0.cm}
\vy\zzz{-0.18cm}
\ee

\zzzz{-0.7cm}\np 
and plugging this prescription into Eqs.~(29) and (31) 
in Ref.~\cite{Holk:2003qa} is engen\nnn-\xyz dering a 
new version (the concrete formula will be presented in 
\sn{4} by\xyz

\zzzz{-1.08cm}\np
(4.2)\2) for  
\vw{\Theta\vvv{\m\m\mu\1\nu}{-1.7}\zzz{-0.31cm}\vvv{(2)}{4.2}} 
on the lattice 
which is gauge invariant, has the correct con\nnn-\xyz

\zzzz{-0.99cm}\np
tinuum limit 
and can\1--\nnn\1in contrast 
to corresponding hitherto developed lattice\xyz  
models\1--\nnn\1simultaneously  fulfil the demands for a uniform 
construction princi\nnn-\xyz

\zzzz{-1.07cm}\np
ple and 
for a Wilson form~\cite{Heller:1984hx,Wilson:1974sk,Zach:1995ni} of
the Hamiltonian component 
\vw{\Theta\vvv{\m4\14}{-2}\zzz{-0.3cm}\vvv{(2)}{5}}\xyz

\zzzz{-0.99cm}\np 
as well as for the structure of 
the trace anomaly~\cite{Adler:zt,Collins:1976yq,Dittrich:yi}\1. With 
(2.5) and\xyz

\zzzz{-0.99cm}\np
(2.7), the simplest lattice versions of all quantities 
\vw{\Theta\vvv{\m\m\mu\1\nu}{-1.7}\zzz{-0.31cm}\vvv{(k)}{4.2}}, \vw{
\Omega\vvv{\4(k)}{5}\zzz{-0.46cm}
\vvv{\alpha\2\beta\2\n\gamma\2\nn\delta\1}{-3.2}
}, and \vw{
{\cal L}^{\zz{\z{\yy\1(\nn\z{\yyy\sf$k$})}}{}}
\mm}\xyz

\zzzz{-0.9cm}\np
are of order one in the plaquette variable or its respective 
tied dual extension\xyz

\zzzz{-1.03cm}\np
while their lattice field strength tensor 
representation is in the non\1-\1dual case\xyz

\zzzz{-0.99cm}\np
of order 
\z{\1"\m$k$\2\nnnn"\m}  
in \vw{\zzzzzz
\zzz{-0.2cm}
\begin{array}{l}
\zzz{0.4cm}\5
\z{\yy\1(\nnn\1$k$\2\mn)}
\zzzz{-0.26cm}\\
{\zz{\z{$\4\cal F\mmmm$}}{\z{\yy$\sim$}}}
\zz{}{\mmmmm\z{\yy\m$\mu\1\nu$}}
\zzzz{-0.cm}
\vvy} as required 
strictly in the continuum limit \z{w\mm.\1r\m.\1t.\mm} \vw{
{\z{\x F}}_{\zz{}{\zzz{-0.12cm}\z{\yyy\zzzzz$\mu\zzz{0.02cm}\nu$}}}
}.\xyz

\zzzz{-0.87cm}\np
We obtain an elegant lattice counterpart for (2.1):

\zzzz{-0.65cm}
\be 
\z{U}\mm\uuu{\2}{\mu\1\n\nu\1}{-3}
\3=\3
\mnn
\sum_{\mbox{\yy $l$$\mmm=\mmm0$}}
^{\zz{\z{\y$\www{2}{-2}$}}{}}
\4
\w
{\2{(\2i\3g\m\mn\vvv{\circ}{-2.7}\2a^{\12}\2)}^{\3\z{\yy$l$}}\2} 
{\zz{}{l\4!}}
\9
{
\ub(
\mmm
\begin{array}{l}
\zzz{0.4cm}\5
\z{\yy \1(\2$l$\2)}
\zzzz{-0.26cm}\\
{\zz{\z{$\4\cal F\mmmm$}}{\z{\yy$\sim$}}}
\zz{}{\mmmmm\z{\yy\m$\mu\1\nu$}}
\zzzz{-0.cm}
\vy
\mmm
\m
\ub)
}
\mm
\zz{\zz{\zz{\zz{\z{\yy$l$}}{}}{}}{}}{}\zzz{-0.1cm}
\z{\7,}
\ee

\zzzz{-0.6cm}\np 
with 
\z{\0\zzz{-0.25cm}
\vww{\zzzzzz
{\m\ub(
\mmm
\begin{array}{l}
\zzz{0.4cm}\5
\z{\yy\1(\nnn\1$0$\1\nnn)}
\zzzz{-0.26cm}\\
{\zz{\z{$\4\cal F\mmmm$}}{\z{\yy$\sim$}}}
\zz{}{\mmmmm\z{\yy\m$\mu\1\nu$}}
\zzzz{-0.cm}
\vy
\mmmm
\ub)}
\mm\mm
\zz{\zz{\zz{\zz{\z{\yy$0$}}{}}{}}{}}{}
\3\www{\equiv}{0.5}\6
\www{\z{\unit{0.06}\mmmm}}{0.05}
}\m.} Although neither (2.8) nor its continuum analogue\xyz

\zzzz{-0.88cm}\np 
(2.1) couple gravity to a Yang\1-Mills theory 
the proximity of both domains\xyz

\zzzz{-1cm}\np
in (2.1) becomes still closer in case of (2.8).

\SEC{Square Roots of Matrices and Semi\nnn-Uni\nnn-\zyx tarity}
\label{Sec_Matrix_Roots}

Let us recall the expressions (2.4) and (2.6) of the preceding 
section\1:

\yzy 

\zzzz{-0.64cm}
\be
\1\mmm
\begin{array}{l}
\zzz{0.4cm}\5
\z{\yy \1(\11\1)}
\zzzz{-0.26cm}\\
{\zz{\z{$\4\cal F\mmmm$}}{\z{\yy$\sim$}}}
\zz{}{\mmmmm\z{\yy\m$\mu\1\nu$}}
\zzzz{-0.cm}
\vy\mmm\m
\2=\5
\vvx 
\z{\xx$\w{1}{\zzz{0.15mm}\z{\y$2\4i\3\kappa$}}$}\zzzz{-0.3mm} 
\vy\vvx
\bigg(\zzzz{0.16cm}\vy\zzz{-0.17cm}
\5  
{\z{U}}_{\mm\zz{}{\z{\yy$\mu\1\nu$}}}
\5-\5
{\z{U}}_{\mm\zz{}{\z{\yy$\mu\1\nu$}}}\zzzz{-0.01cm}
\zzz{-0.11cm}  
\vx
\sfdagger{0}
\zzzz{0.5cm} 
\vy\zzz{-0.19cm} 
\0\vx
\bigg)\zzzz{0.16cm}\vvy
\z{\7,} 
\ee

\zzzz{-0.64cm}
\be
\1\mmm
\begin{array}{l}
\zzz{0.4cm}\5
\z{\yy \1(\12\1)}
\zzzz{-0.26cm}\\
{\zz{\z{$\4\cal F\mmmm$}}{\z{\yy$\sim$}}}
\zz{}{\mmmmm\z{\yy\m$\mu\1\nu$}}
\zzzz{-0.cm}
\vy\mmm\m
\2=\5
\vvx
\z{\xx$\w{1}{\z{\y$i\2\kappa$}}$}\zzzz{-0.3mm} 
\vy\vvx
\Bigg(\zzzz{0.2cm}\vy\zzz{-0.17cm}
\5
\z{\zzzzz$\zzzzzz\sqrt{\z{\unboldmath$\zzzzzz
\zzz{-0.08cm}\vx
\zzzz{-0.48cm}\\
{\z{U}}_{\mm\zz{}{\z{\yy$\mu\1\nu$}}}\zzzz{-0.01cm}
\vy\zzz{-0.cm}
$}}$}
\zzz{-0.21cm}\vx
\z{\yyyy\zzzzz$|\zzz{-0.098cm}|$}\zzzz{0.595cm}    
\vy\zzz{-0.21cm}\vx
\z{\y\tt main}\zzzz{0.51cm}  
\vy\zzz{-0.17cm}
\6
-
\4\zzz{0.1mm}
\vvx
\nn\m 
\zzz{0.2mm}
\zzz{-0.13cm}\vx
\zzzz{-0.6cm}\\
\z{\zzzzz$\zzzzzz\sqrt{\z{\uuuuu$\zzzzzz
\zzz{-0.08cm}\vx
\zzzz{-0.48cm}\\
{\z{U}}_{\mm\zz{}{\z{\yy$\mu\1\nu$}}}\zzzz{-0.01cm}
\vy\zzz{-0.cm}
$}}$}
\zzz{-0.21cm}\vx
\z{\yyyy\zzzzz$|\zzz{-0.098cm}|$}\zzzz{0.595cm}    
\vy\zzz{-0.21cm}\vx
\z{\y\tt main}\zzzz{0.51cm}  
\vy\zzz{-0.17cm}
\zzzz{-0.2cm}
\vy\zzz{-0.11cm}
\m 
\zzzz{0.1cm}
\vvy
\zzz{-0.08cm}\vx
\sfdagger{0}
\m 
\zzzz{0.7cm}
\vy\zzz{-0.19cm}
\2\vx
\Bigg)\zzzz{0.2cm}\vvy
\z{\7,}
\ee

\zzzz{-0.6cm}
\np 
with \vww{
\kappa\1\defin\1
g\zzz{-0.03cm}\vvv{\circ}{-2.7}\mnnn(a)\4
a^{\12}
} being different from the former 
\vw{\kappa} in Ref.~\cite{Holk:2003qa}\1. On the\xyz

\zzzz{-0.98cm}\np 
other hand, the standard field strength tensor on the 
lattice, \vw{\zzzzzz
\vvx{\zz{\z{$\4\cal F\mmmm$}}{\z{\yy$\sim$}}}
\zz{}{\mmmmm\z{\yy\m$\mu\1\nu$}}\zzzz{-0.2cm}
\vvy\2}, is 
defined\xyz by

\zzzz{-1.1cm}
\be
{\z{U}}_{\mm\zz{}{\z{\yy$\mu\1\nu$}}}
\2=\5
\zzz{-1.6cm}
\vx
\zzz{1.7cm}
i\3\kappa
\mmmmm\m
\vx
{\zz{\z{\y$\4\cal F\mmmm$}}{\z{\yyy$\sim$}}}
\zz{}{\mmmmm\z{\yyy\m$\mu\1\nu$}}
\zzzz{-0.2cm}
\vy
\zzzz{-0.1cm}
\\
\z{\xx$e$}
\zzzz{0.4cm}
\vvy
\zzz{-0.15cm}
\0 
\ee

\zzzz{-0.8cm}
\np 
implicitly. (3.1) up to (3.3) can be put together, yielding

\zzzz{-0.9cm}
\be
\0
\zzz{-0.18cm}
\begin{array}{l}
\zzz{0.4cm}\5
\z{\yy \1(\2$l$\2)}
\zzzz{-0.26cm}\\
{\zz{\z{$\4\cal F\mmmm$}}{\z{\yy$\sim$}}}
\zz{}{\mmmmm\z{\yy\m$\mu\1\nu$}}
\zzzz{-0.cm}
\vy\mmm\m
\m=\3
\zzz{-0.18cm}\vx
\z{\xx$\w{l}{\zz{\z{\y$\kappa$}}{}}$}
\vy\zzz{-0.18cm}
\6
\z{\sl sin}
\1\zzz{-0.1mm}\z{\zzzzz$\bigg($}\3
\zzz{-0.18cm}\vx
\z{\xx$\w{\kappa}{{\z{\y$l$}}{}}$}\zzzz{-0.11cm}
\vy\zzz{-0.18cm}\6
\z{mod\m\mnn} 
\2\Big(\3
\mmm\mmmmm\vx
{\zz{\z{$\4\cal F\mmmm$}}{\z{\yy$\sim$}}}
\zz{}{\mmmmm\z{\yy\m$\mu\1\nu$}}
\zzzz{-0.2cm}
\vy\mmm\mmmmm
\5,\2
\z{\tt period}
\4\Big)\zzz{-0.6mm}\z{\zzzzz$\bigg)$}\2
\zzz{0.7cm}
\forall\6\4\6\4\6l\9\in\6
\{\2\z{$1\2,\12\2\}$}
\z{\7.}
\ee

\zzzz{-0.56cm}\np 
(3.4) has to be understood as an expression which is purely symbolic  
in case\xyz of the underlying gauge group being nonabelian and is then 
motivated by its\xyz own reinterpretation as a continuation of the abelian 
particular case where\xyz the most general gauge group is \z{\n\0U(1)\n} 
and (3.4) has the precise significations\xyz

\zzzz{-1.08cm}
\be
\0
\zzz{-5.22cm}
\0
\zzz{-0.18cm}\begin{array}{l}
\zzz{0.4cm}\5
\z{\yy \1(\11\1)}
\zzzz{-0.26cm}\\
{\zz{\z{$\4\cal F\mmmm$}}{\z{\yy$\sim$}}}
\zz{}{\mmmmm\z{\yy\m$\mu\1\nu$}}
\zzzz{-0.cm}
\vy\mmm\m
\2=\5
\zzz{-0.18cm}\vx
\z{\xx$\w{1}{\zz{\z{\y$\kappa$}}{}}$}
\vy\zzz{-0.18cm}
\7
\z{\sl sin}\5
{\vartheta}_{\mmm\zzz{0.1mm}\zz{}{\mu\1\nu}}
\3=\5 
\zzz{-0.18cm}\vx
\z{\xx$\w{1}{\zz{\z{\y$\kappa$}}{}}$}
\vy\zzz{-0.18cm}
\7
\z{\sl sin}\5
\zzz{-0.13cm}\z{\zzzzz$\bar{\z{\unboldmath$\zzz{0.13cm}
{\vartheta}_{\mmm\zzz{0.1mm}\zz{}{\mu\1\nu}}    
\zzz{-0.13cm}$}}$}\zzz{0.15cm}
\z{\7,}
\ee

\zzzz{-1.1cm} 
\be
\0
\zzz{-0.18cm}
\begin{array}{l}
\zzz{0.4cm}\5
\z{\yy \1(\12\1)}
\zzzz{-0.26cm}\\
{\zz{\z{$\4\cal F\mmmm$}}{\z{\yy$\sim$}}}
\zz{}{\mmmmm\z{\yy\m$\mu\1\nu$}}
\zzzz{-0.cm}
\vy\mmm\m
\2=\5
\zzz{-0.18cm}\vx
\z{\xx$\w{2}{\zz{\z{\y$\kappa$}}{}}$}
\vy\zzz{-0.18cm}
\6
\z{\sl sin}   
\5
\lim_{\z{\yyy\zzzzz$\zzzzzz\varepsilon$}
\1\rightarrow\1\zzz{0.15mm}0
}\1
\zzz{-0.18cm}\vx
\z{\xx$\w{\zzz{0.05mm}
\z{\y$
\zzz{-0.18cm}\vx
\z{\y mod}\0(\2 
{\vartheta}_{\mmm\zzz{0.1mm}\zz{}{\mu\1\nu}}
\m-\mm
\z{\y\zzzzz$|$}\1\z{\y\zzzzz$\varepsilon$}\0\z{\y\zzzzz$|$}
\3\zzz{0.1mm}\z{\y\bf ,}\52\2\pi\1)\m-\pi
\zzzz{0.04cm}\vy\zzz{-0.18cm}
$}
\zzz{0.1mm}}{{\z{\y$
\zzz{-0.18cm}\vx\zzzz{-0.53cm}\\
2
\vy\zzz{-0.18cm}
$}}{}}$}\zzzz{0.07cm}
\vy\zzz{-0.18cm}
\5=\5
\zzz{-0.18cm}\vx
\z{\xx$\w{2}{\zz{\z{\y$\kappa$}}{}}$}
\vy\zzz{-0.18cm}
\6
\z{\sl sin}
\5
\zzz{-0.18cm}\vx
\z{\xx$\w{
\z{\y$
\zzz{-0.18cm}\vx
\zzz{-0.13cm}\z{\zzzzz$\bar{\z{\unboldmath$\zzz{0.13cm}
{\vartheta}_{\mmm\zzz{0.1mm}\zz{}{\mu\1\nu}}    
\zzz{-0.13cm}$}}$}\zzz{0.13cm}
\zzzz{0.04cm}\vy\zzz{-0.18cm}
$}
}{{\z{\y$
\zzz{-0.18cm}\vx\zzzz{-0.53cm}\\
2
\vy\zzz{-0.18cm}
$}}{}}$}\zzzz{0.07cm}
\vvy
\z{\7,}
\ee

\zzzz{-0.58cm}\np
using the \z{\n\0U(1)\n} gauge field decomposition\zyx

\zzzz{-0.85cm}
\np\vww{\zzzzzz
{\vartheta}_{\mmm\nn\zz{}{\mu\1\nu}}
\1\defin\1
\kappa\4
\mmm\mmmmm\vx
{\zz{\z{$\4\cal F\mmmm$}}{\z{\yy$\sim$}}}
\zz{}{\mmmmm\z{\yy\m$\mu\1\nu$}}
\zzzz{-0.2cm}
\vy\mmm\mmmmm
\3=\3
\sum_{l\1\zzz{0.1mm}=\11}^{4}\4
\vartheta\zzz{-0.066cm}
\uuu{\1\m}{(}{-3.4}
\uuu{\1}{\2l\2}{-3.9}
\uuu{\1\m}{)}{-3.4}
\6\4\5\in\8\4
\z{\bf ]}\m-\m4\2\pi\3,\14\2\pi\4
\z{\bf ]}
} in 
the exponent of the plaquette\xyz

\zzzz{-0.61cm}\np
variable and the improvement of this representation via\zyx

\zzzz{-0.86cm}
\np\vww{\zzzzzz
{\vartheta}_{\mmm\zzz{0.1mm}\zz{}{\mu\1\nu}}
\3=\2
\z{\zzzzz$\bar{\z{\unboldmath$\zzz{0.13cm} 
{\vartheta}_{\mmm\zzz{0.1mm}\zz{}{\mu\1\nu}}    
\zzz{-0.13cm}$}}$}
\7+\3
2\3n\2\pi
\z{\7,}\zzz{0.2cm}
n\9\in\6
\{\1\z{$-\12\3,-1\2,\20\3,1\2,\12\3\}$}
\z{\7,}\zzz{0.2cm}
\z{with}\9
\z{\zzzzz$\bar{\z{\unboldmath$\zzz{0.13cm} 
{\vartheta}_{\mmm\zzz{0.1mm}\zz{}{\mu\1\nu}}    
\zzz{-0.13cm}$}}$}
\zzz{0.16cm}
\5\in\8
\z{\bf ]}\m-\m\pi\3,\1\pi\4
\z{\bf ]}
\z{\7.}
}

\zzzz{0.2cm}
We have already given a prescription for the calculation of the 
matrix\xyz  main\1-\nnn\0value roots in the last section, 
regarding the case of diagonalizable 
ma\1-\xyz trices. We will not discuss the scope 
for an eventual generalization to other\xyz square matrices here 
where the product of the main root of the diagonalizable\xyz part 
and a finite binomial series for the nilpotent matrix\1-\nnn\0valued 
remnant fac\1-\xyz tor (\1up to the highest multiplicity of any original 
eigenvalue minus one\1) has to be taken into account. We just remark 
that the self\nnn\1-\1consistency of 
the main\xyz

\zzzz{-0.99cm}\np
root of a matrix \vw{M} in case of 
diagonalizability \vww{\zzzzzz
M
\2=\2
A\5
D\3
A^{\1-\11}
} (\nn\0with \vw{D} 
be\2-\xyz

\zzzz{-0.93cm}\np 
ing a diagonal matrix\nnnn) can 
be illustrated 
via \vww{\zzzzzz
\z{\yy$\zzzzzz
\0
\uuu{\5}{\z{(}}{0.9}
\1
\www{
\z{\y$\zzzzzz\sqrt{\nnnn\z{\yy$\zzzzzz 
\vvx\zzzz{-0.51cm}\\
A\5
D\3
A
\uuu{\0}{\mmm-\mm1}{2.8}
\\\zzzz{-0.69cm}\vvy
$}\6}$}
}{1.3}
\5\nnn
\uuu{\5}{\z{)}}{0.9}
\uuu{\1}{\12}{4.9}
\3\nnn=\4\mn
\uuu{\4}{\z{(}}{1}
\4
A\5
\sqrt{\nnnn\0D\6}
\4
A
\uuu{\0}{\mmm-\mm1}{2.8}
\4
\uuu{\4}{\z{)}}{1}
\uuu{\0}{\22}{4.6}
\0
\0
$}
}\xyz

\zzzz{-1.02cm}\np
and the implementation of unicity 
by introducing the main\1-\nnn\0value prescrip\nn\1-\xyz tion 
for the involved square roots. It is clear that any element of 
any unitary\xyz group is diagonalizable. Besides, we would like to 
indicate following subtle\2-\xyz

\yzy 

ty\1: in case of the special 
unitarian gauge group \z{\n\0SU(2)\n}, e.\1g.\1, the main\1-\1root\xyz  
operation will exceed the group if and only if the tackled group 
element is\xyz unequal to the negative of the identity matrix. But 
without exception any\xyz main (\nnn\0principal\nnnn) root of an 
element of any orthogonal, special orthogonal or\xyz special unitary 
group lies in the surrounding, merely unitarian supergroup.

Restoration of rotational invariance on the lattice can be 
achieved by the substitution of the plaquette variable

\zzzz{-0.75cm}   
\be
\z{U}\uuu{\2}{\mm\mnn\mu\1\nnnn\nu}{-3}\1(n)
\3\equiv\3
\z{U}
\1(\n\0n\2\n;\mu\2\nnn,\nu\1)
\ee

\zzzz{-0.6cm}
\np 
by the averaging

\zzzz{-0.65cm}   
\wx
\0
\zzz{-1.35cm}
\z{V}\uuu{\2}{\mm\mnnnnn\mu\1\nnnn\nu}{-3}\1(n)
\3\defin\3
\vvx\z{\xx$\w{1}{4}$}\vvy
\5
\uuu{\7\bf}{[}{-2}
\6
\z{U}
\1(\n\0n\2\n;\mu\2\nnn,\nu\1)
\4\nn+\4
\z{U}
\1(\n\0n\2\n;-\mu\2\nnn,\nu\1)
\4\nn+
\wy

\zzzz{-1cm}
\be
\0
\zzz{1.35cm}
+\4
\z{U}
\1(\n\0n\2\n;-\mu\2\nnn,\mnnnnnn-\nnnnnn\nu\1)
\4\nn+\4
\z{U}
\1(\n\0n\2\n;\mu\2\nnn,\mnnnnnn-\nnnnnn\nu\1)
\5
\uuu{\7\bf}{]}{-2}
\ee

\zzzz{-0.35cm}
\np 
of all four plaquettes touching the 
\z{\four\1space} lattice site \z{\1"\mnnn$n$\2"\m} in 
the \z{\2$\mu$\2-\1$\nu$} plane.\xyz For 
reasons of gauge invariance, the respective circulation has to 
start and\xyz to end at the same \z{\1"\mnnn$n$\2"\m}, preserving 
one scheme of orientation, being coun\nnn-\xyz terclockwise 
according to usual convention. Let us call 
a \z{\12\2-\1by\2-\12\1} matrix 
\vw{\www{\z{\x\tt U}}{-0.45}}\xyz 
semi\1-\nnn\0unitary here if

\zzzz{-0.78cm}
\be
{\www{\z{\x\tt U}}{-0.45}}
\7
{\www{\z{\x\tt U}}{-0.45}}
\1
\uuu{\2\m}{\dagger}{4}
\4=\3
\6\6\6\www{\unit{0.06}}{-0.5}\7
\6
\mm\z{\sl\1a\mn\0b\nnn\0s}\6 
\z{\sl det}\9
{\www{\z{\x\tt U}}{-0.45}}
\ee

\zzzz{-0.58cm}
\np 
is valid and 
"\m\0special semi\1-\nnn\0unitary\2" 
if both (3.9) and

\zzzz{-0.68cm}
\be
\z{\sl det}\9
{\www{\z{\x\tt U}}{-0.45}}
\6=\5
\z{\sl Re}\6
\z{\sl det}\9
{\www{\z{\x\tt U}}{-0.45}}
\9
\uuu{\1\m}{\ge}{1}
\80
\ee

\zzzz{-0.48cm}
\np 
are fulfilled. Hence the special case of

\zzzz{-0.71cm}
\be
\z{\sl\1a\mn\0b\nnn\0s}\6 
\z{\sl det}\9
{\www{\z{\x\tt U}}{-0.45}}
\4=\3
1
\ee

\zzzz{-0.5cm}\np
complementally changes
semi\1-\nnn\0unitarity 
into unitarity and 
special semi\1-\nnn\0uni\1-\xyz

\zzzz{-1cm}\np
tarity
into special unitarity. If \vw{
\z{U}\uuu{\2}{\mm\mnn\mu\1\nnnn\nu}{-3}\1(n)
} in (3.7) is 
an element of the gauge\xyz

\zzzz{-0.88cm}\np 
group 
\z{\n\0SU(2)\n} 
then \vw{
\z{V}\uuu{\2}{\mm\mnnnnn\mu\1\nnnn\nu}{-3}\1(n)
} in (3.8) is 
special semi\1-\nnn\0unitary.

\zzzz{0.3cm}
The main root of such a \vw{
\z{V}\uuu{\2}{\mm\mnnnnn\mu\1\nnnn\nu}{-3}\1(n)
} is just 
semi\1-\nnn\0unitary 
if \vww{
\z{V}
\1\defin\1
\z{V}\uuu{\2}{\mm\mnnnnn\mu\1\nnnn\nu}{-3}\1(n)
} is 
a\xyz

\zzzz{-0.88cm}\np 
(\1stretching\1/scaling down\nn) point\1-\1symmetry reflection

\zzzz{-0.66cm}
\be
\z{V}
\7=\2-\2
\mroot{
(\1\z{\sl det}\7\z{V}\3\nn)
}{0}
\6\2
\uuu{\5\m}{\cdot}{-0.5}
\2\9\6\6\www{\unit{0.06}}{-0.5}\7
\ee

\zzzz{-0.68cm}
\np 
(\1the main root is here equal to the habitual principal value of 
the square\xyz root, 
\z{v\mm.\1s\n.)} 
and special semi\1-\nnn\0unitary 
else. Eq.~(3.12) is 
the \vww{
\z{\zzzzz$\alpha$}\nnn
(\1\mn\z{V}\1)
=\mmm-\mm1
}~case\xyz of the 
"\m\0semi\1-\nnn\0unitarity signature\2"

\samepage{
\zzzz{-0.75cm}
\wx
\z{\zzzzz$\alpha$}\nnn
(\1\mn\z{V}\1)
\5\defin\6
2\6
\z{sgn}\5\z{\sl det}
\3
\uuu{\5}{\z{(}}{0}
\5\nnn
\z{V}
\3+\4
\mroot{\5\z{\sl det}\7\z{V}\8}{0}
\8
\uuu{\5\m}{\cdot}{-0.5}
\4\9\6\6\www{\unit{0.06}}{-0.5}\7
\6
\uuu{\5}{\z{)}}{0}
\1-\41
\zzz{0.6cm}\in\3\8\z{$\{\mm-\mmm1\4,\mmm+\mm1\2\}$}
\z{\7,}
\wy

\zzzz{-1.15cm}
\be\ee
}

\yzy 

\zzzz{-0.8cm}
\np 
erecting the evaluation criterion

\zzzz{-0.6cm}
\be
\uuu{\m}{\alpha}{0}
\mn\2(\n\4\rho\n\3,\nnn\1\sigma\nn\5\z{\bf ;}\n\6\xi\n\3,\nnn\1\eta\3\nnnn)  
\4\defin\4
\vvx\z{\xx$\w{1}{2}$}\vvy
\4
\uuu{\6}{\z{(}}{-1}
\1\nnn
\uuu{\m}{\alpha}{0}\2
\uuu{\5}{\z{(}}{0}
\1
\z{V}\uuu{\2}{\mmm\mnnnnn\rho\1\nnnn\sigma}{-3}
\2
\uuu{\5}{\z{)}}{0}
\nnn+\1
\uuu{\m}{\alpha}{0}\2
\uuu{\5}{\z{(}}{0}
\1
\z{V}\uuu{\2}{\m\mnnnnn\xi\1\nnnn\eta}{-3}
\2
\uuu{\5}{\z{)}}{0}
\1
\uuu{\6}{\z{)}}{-1}
\ee

\zzzz{-0.5cm}
\np 
for the kernel

\zzzz{-0.6cm} 
\be
\zzz{-2cm}
\uuu{\5\m}{\langle}{0}
\6
\uuu{\5}{\rho}{0}\5
\uuu{\5}{\sigma}{0}\5
\uuu{\5}{\xi}{0}\5
\uuu{\5}{\eta}{0}
\6
\uuu{\5\m}{\rangle}{0}
\3\defin\4\nnn
\z{\sl trace}
\0\mnn
\ux(
\vx\zzzz{-0.7cm}\\
\mm\mnnn
\mmmm\mmmm\begin{array}{l}\zzz{0.4cm}\5\z{\yy\1(\12\1)}
\zzzz{-0.26cm}\\{\zz{\z{$\4\cal F\mmmm$}}{\z{\yy$\sim$}}}
\zz{}{\mmmmm\z{\yy\m$\rho\1\sigma$}}\zzzz{-0.cm}\vy\mmm\mmm
\6
\mmmm\mmmm\begin{array}{l}\zzz{0.4cm}\5\z{\yy\1(\12\1)}
\zzzz{-0.26cm}\\{\zz{\z{$\4\cal F\mmmm$}}{\z{\yy$\sim$}}}
\zz{}{\mmmmm\z{\yy\m$\xi\1\nnn\eta$}}\zzzz{-0.cm}\vy\mmm\mmm
\mmmm
\\\zzzz{-0.7cm}\vy
\uy)
\z{\7,}
\ee

\zzzz{-0.5cm}
\np 
which is proportional 
to \vw{
\Omega\vvv{\4(2)}{5}\zzz{-0.46cm}
\vvv{\mn\rho\2\sigma\2\mn\xi\2\mn\eta\1}{-2.8}
} in \sn{2} if the 
limit \vww{a\2\ra\2\nn0} is investigated.

\zzzz{0.2cm}
There are two possible refinements 
for \vw{\zzzzzz
\zzz{-0.2cm}
\begin{array}{l}
\zzz{0.4cm}\5
\z{\yy\1(\2$l$\2)}
\zzzz{-0.26cm}\\
{\zz{\z{$\4\cal F\mmmm$}}{\z{\yy$\sim$}}}
\zz{}{\mmmmm\z{\yy\m$\mu\1\nu$}}
\zzzz{-0.cm}
\vvy} in (3.1) and (3.2). First, 
the\xyz

\zzzz{-0.9cm}\np 
steps (3.1) and (3.2) may admix 
\z{\n\0U(N)\n} generators 
to \vw{\zzzzzz
\zzz{-0.2cm}
\begin{array}{l}
\zzz{0.4cm}\5
\z{\yy\1(\2$l$\2)}
\zzzz{-0.26cm}\\
{\zz{\z{$\4\cal F\mmmm$}}{\z{\yy$\sim$}}}
\zz{}{\mmmmm\z{\yy\m$\mu\1\nu$}}
\zzzz{-0.cm}
\vvy} which 
are outside\xyz

\zzzz{-0.8cm}\np
of the original Lie algebra 
spanning \vw{\zzzzzz
\vvx{\zz{\z{$\4\cal F\mmmm$}}{\z{\yy$\sim$}}}
\zz{}{\mmmmm\z{\yy\m$\mu\1\nu$}}\zzzz{-0.2cm}
\vvy\2} in (3.3) 
if this original Lie algebra\xyz

\zzzz{-0.8cm}\np
refers to the 
non\1-\1trivial subgroups
\z{\n\0O\nn(N)\n,} 
\z{\n\0SO\nn(N)\n,} or 
\z{\n\0SU(N)\n.} Such 
artefact\xyz

\zzzz{-1cm}\np
generators can be removed by the reprojection

\zzzz{-0.6cm}
\samepage{
\wx
\zzz{-0.2cm}\begin{array}{l}\zzz{0.4cm}\5\z{\yy \1(\2$l$\2)}
\zzzz{-0.26cm}\\{\zz{\z{$\4\cal F\mmmm$}}{\z{\yy$\sim$}}}
\zz{}{\mmmmm\z{\yy\m$\mu\1\nu$}}\zzzz{-0.cm}\vy\zzz{-0.2cm}
\zzz{0.8cm}
\bp{1}{16}{1}
\llp
\put(0,1.08){\vector(1,0){16}}
\ep
\zzz{0.8cm}
%
\zzz{-0.2cm}\begin{array}{l}\zzz{0.4cm}\5\z{\yy \1(\2$l$\2)}
\zzzz{-0.26cm}\\{\zz{\z{$\4\cal F\mmmm$}}{\z{\yy$\sim$}}}
\zz{}{\mmmmm\z{\yy\m$\mu\1\nu$}}\zzzz{-0.cm}\vy\zzz{-0.2cm}
\zzz{-0.033cm}
\uuu{\5\tt}{\z{[}}{2.8}\m
\uuu{\2\sf}{\z{R}}{3.5}\m
\uuu{\5\tt}{\z{]}}{2.8}
\3
\defin
\5
2\6 
\z{\sl trace}
\2
\www{\z{\xxx(}}{-0.2}
\nnn\2
\zzz{-0.2cm}\begin{array}{l}\zzz{0.4cm}\5\z{\yy \1(\2$l$\2)}
\zzzz{-0.26cm}\\{\zz{\z{$\4\cal F\mmmm$}}{\z{\yy$\sim$}}}
\zz{}{\mmmmm\z{\yy\m$\mu\1\nu$}}\zzzz{-0.cm}\vy\zzz{-0.2cm}
\6
{\zzz{-0.13cm}\vx\z{\xx\zzzzz$\hat{\z{\x\uuuuu$\zzzzzz
\vvx\zzzz{-1.3cm}\\
\zzz{-0.025cm}\tau\zzz{0.025cm}\zzzz{-0.58cm}\vy\zzz{-0.18cm}
$}}$}\zzzz{-0.08cm}\vy\zzz{-0.15cm}}
\uuu{\1}{A}{3}
\n\2
\www{\z{\xxx)}}{-0.2} 
\4
{\zzz{-0.13cm}\vx\z{\xx\zzzzz$\hat{\z{\x\uuuuu$\zzzzzz
\vvx\zzzz{-1.3cm}\\
\zzz{-0.025cm}\tau\zzz{0.025cm}\zzzz{-0.58cm}\vy\zzz{-0.18cm}
$}}$}\zzzz{-0.08cm}\vy\zzz{-0.3cm}}
\uuu{\1}{A}{-3.5}\2
\wy

\zzzz{-0.84cm}
\be
\0
\zzz{0.38cm}
\forall\6\4\6l\6\in\4
\{\2\z{$1\2,\12\2\}$}
\ee
}

\zzzz{-0.88cm}\np 
onto the original Lie algebra. Second, rotational invariance on the 
lattice can be implemented by

\zzzz{-0.6cm}
\samepage{
\wx
\mmmm\mmmm\begin{array}{l}\zzz{0.4cm}\5\z{\yy\1(\2$l$\2)}
\zzzz{-0.26cm}\\{\zz{\z{$\4\cal F\mmmm$}}{\z{\yy$\sim$}}}
\zz{}{\mmmmm\z{\yy\m$\mu\1\nu$}}\zzzz{-0.cm}\vy\mmm\mmm
\mnn
\uuu{\7}{\z{(}}{-2}
\2
\z{U}
\uuu{\2}{\mm\mnn\mu\1\nnnn\nu}{-3}
\1(n)     
\uuu{\7}{\z{)}}{-2}
\zzz{0.8cm}
\bp{1}{16}{1}
\llp
\put(0,1.08){\vector(1,0){16}}
\ep
\zzz{0.8cm}
%
\mmmm\mmmm\begin{array}{l}\zzz{0.4cm}\5\z{\yy\1(\2$l$\2)}
\zzzz{-0.26cm}\\{\zz{\z{$\4\cal F\mmmm$}}{\z{\yy$\sim$}}}
\zz{}{\mmmmm\z{\yy\m$\mu\1\nu$}}\zzzz{-0.cm}\vy\mmm\mmm
\mnn
\uuu{\7}{\z{(}}{-2}
\2
\z{U}
\uuu{\2}{\mm\mnn\mu\1\nnnn\nu}{-3}
\1(n)     
\zzz{0.2cm}
\uuu{\m}{\rra}{0}
\zzz{0.2cm}
\z{V}
\uuu{\2}{\mm\mnnnnn\mu\1\nnnn\nu}{-3}
\1(n)
\2\mnnnn
\uuu{\7}{\z{)}}{-2}
\wy

\zzzz{-0.82cm}
\be
\0
\zzz{3.78cm}
\forall\6\4\6l\6\in\4
\{\2\z{$1\2,\12\2\}$}
\z{\7,}
\ee
}

\zzzz{-0.88cm}\np
according to (3.7) and (3.8). If (3.16) is applied definitely and 
(3.17) may\xyz

\zzzz{-1cm}\np
be supplemented optionally then 
the\nnn\1--\1admittedly not finally fixed\1--\nnn\1resulting\xyz

\zzzz{-1cm}\np
refinement 
shall be denoted 
by \z{\vw{\zzzzzz
\zzz{-0.2cm}
\begin{array}{l}
\zzz{0.4cm}\5
\z{\yy\1(\2$l$\2)}
\zzzz{-0.26cm}\\
{\zz{\z{$\4\cal F\mmmm$}}{\z{\yy$\sim$}}}
\zz{}{\mmmmm\z{\yy\m$\mu\1\nu$}}
\zzzz{-0.cm}
\vvy
\zzz{0.4mm}\uuu{\5}{\prime}{3.66}\n
}\m.}

\zzzz{0.2cm} 
The more general evaluation of (3.14) and (3.15) is rendered possible 
if\xyz

\zzzz{-0.92cm}\np
the 
\z{\14\2-\nnnn\0plaquettes} mixing 
in (3.8) and (3.17) 
is switched on\zyx

\zzzz{-0.92cm}\np
\z{(\vww{
\uuu{\4}{\sqrt{\rule{0cm}{0.5cm}\zzz{0.93cm}}}{-0.1}
\2=\2
\mroot{\rule{0cm}{0.1cm}\zzz{0.66cm}}{0}
}):}

\zzzz{-1cm}
\samepage{
\wx 
\mnn
\uuu{\5\m}{\langle}{0}
\6
\uuu{\5}{\rho}{0}\5
\uuu{\5}{\sigma}{0}\5
\uuu{\5}{\xi}{0}\5
\uuu{\5}{\eta}{0}
\6
\uuu{\5\m}{\rangle}{0}
\2
=
\3
\w{4}{\vvx\zzzz{-0.6cm}\\{\kappa}\vvv{\nnnn\12}{4}\vvy}
\8
\w{
\z{\sl Im}\5
\www{\z{V}}{-0.05}
\uuu{\2}{\mm\mnnnnn\rho\1\nnnn\sigma}{-3.05}\zzz{-0.3cm}
\vvv{1\21}{5}
\8
\z{\sl Im}\5
\www{\z{V}}{-0.05}
\uuu{\2}{\mm\mnnnnn\1\xi\1\nnnn\eta}{-3.05}\zzz{-0.3cm}
\vvv{1\21}{5}
\6+\3
\z{\sl Re}\5
\www{\z{V}}{-0.05}
\uuu{\2}{\mm\mnnnnn\rho\1\nnnn\sigma}{-3.05}\zzz{-0.3cm}
\vvv{1\22}{5}
\8
\z{\sl Re}\5
\www{\z{V}}{-0.05}
\uuu{\2}{\mm\mnnnnn\1\xi\1\nnnn\eta}{-3.05}\zzz{-0.3cm}
\vvv{1\22}{5}
\6+\3
\z{\sl Im}\5
\www{\z{V}}{-0.05}
\uuu{\2}{\mm\mnnnnn\rho\1\nnnn\sigma}{-3.05}\zzz{-0.3cm}
\vvv{1\22}{5}
\8
\z{\sl Im}\5
\www{\z{V}}{-0.05}
\uuu{\2}{\mm\mnnnnn\1\xi\1\nnnn\eta}{-3.05}\zzz{-0.3cm}
\vvv{1\22}{5}
}{\6
\vvx\zzzz{-0.36cm}\\
\sqrt{\7
\vvx\zzzz{-0.66cm}\\
\z{\sl Re}\5
\www{\z{V}}{-0.05}
\uuu{\2}{\mm\mnnnnn\rho\1\nnnn\sigma}{-3.05}\zzz{-0.3cm}
\vvv{1\21}{5}
\mnnn\7+\4
\sqrt{\5
\z{\sl det}\7
\www{\z{V}}{-0.05}
\uuu{\2}{\mm\mnnnnn\rho\1\nnnn\sigma}{-3.05}
\6\2}
\zzzz{-0.2cm}
\vvy
\8\3}
\8
\sqrt{\7
\vvx\zzzz{-0.66cm}\\
\z{\sl Re}\5
\www{\z{V}}{-0.05}
\uuu{\2}{\mm\mnnnnn\1\xi\1\nnnn\eta}{-3.05}\zzz{-0.3cm}
\vvv{1\21}{5}
\mnnn\7+\4
\sqrt{\5
\z{\sl det}\7
\www{\z{V}}{-0.05}
\uuu{\2}{\mm\mnnnnn\1\xi\1\nnnn\eta}{-3.05}
\6\2}
\zzzz{-0.2cm}
\vvy
\8\3}
\vvy
\6}
\wy

\zzzz{-0.6cm}
\be
\0
\zzz{2.055cm}
\z{
for 
\zzz{0.2cm}
$\zzzzzz
\uuu{\m}{\alpha}{0}
\mn\2(\n\4\rho\n\3,\nnn\1\sigma\nn\5\z{\bf ;}\n\6\xi\n\3,\nnn\1\eta\3\nnnn)
\4=\1+\mnn\11
$}
\ee
}

\yzy 

\zzzz{-0.8cm}   
\be
\0
\zzz{-0.455cm}
\zzz{-2.028cm}
\uuu{\5\m}{\langle}{0}
\6
\uuu{\5}{\rho}{0}\5
\uuu{\5}{\sigma}{0}\5
\uuu{\5}{\xi}{0}\5
\uuu{\5}{\eta}{0}
\6
\uuu{\5\m}{\rangle}{0}
\5=\5
0
\zzz{1cm}
\z{
for 
\zzz{0.2cm}
$\zzzzzz
\uuu{\m}{\alpha}{0}
\mn\2(\n\4\rho\n\3,\nnn\1\sigma\nn\5\z{\bf ;}\n\6\xi\n\3,\nnn\1\eta\3\nnnn)
\4=\40
$}
\ee

\zzzz{-0.4cm}
\samepage{
\wx
\uuu{\5\m}{\langle}{0}
\6
\uuu{\5}{\rho}{0}\5
\uuu{\5}{\sigma}{0}\5
\uuu{\5}{\xi}{0}\5
\uuu{\5}{\eta}{0}
\6
\uuu{\5\m}{\rangle}{0}
\5
=
\6
\w{8}{\vvx\zzzz{-0.6cm}\\{\kappa}\vvv{\nnnn\12}{4}\vvy}
\8\4
\sqrt{\4
\vvx\zzzz{-0.6cm}\\
\uuu{\6}{\z{(}}{-1}
\3\1
\www{\z{V}}{-0.05}
\uuu{\2}{\mm\mnnnnn\rho\1\nnnn\sigma}{-3.05}
\nn\6\1\uuu{\5}{\cdot}{0}\nnnn
\5\nnnn\1
\www{\z{V}}{-0.05}
\uuu{\2}{\mm\mnnnnn\1\xi\1\nnnn\eta}{-3.05}
\4\nn\1
\uuu{\6}{\z{)}}{-1}
\1
\vvv{1\21}{5}
\\\zzzz{-0.7cm}\vvy
\8}
\wy

\zzzz{-1.1cm}
\be
\0
\zzz{1.97cm}
\z{
for 
\zzz{0.2cm}
$\zzzzzz
\uuu{\m}{\alpha}{0}
\mn\2(\n\4\rho\n\3,\nnn\1\sigma\nn\5\z{\bf ;}\n\6\xi\n\3,\nnn\1\eta\3\nnnn)
\4=\m-\mnn\m1
$}
\ee
}

\zzzz{-0.6cm}\np 
The substitution (3.16) will leave (3.18) and (3.19) unchanged while 
(3.20) then becomes

\zzzz{-0.6cm}
\be
\zzz{-0.455cm}
\zzz{-2cm}
\uuu{\5\m}{\langle}{0}
\6
\uuu{\5}{\rho}{0}\5
\uuu{\5}{\sigma}{0}\5
\uuu{\5}{\xi}{0}\5
\uuu{\5}{\eta}{0}
\6
\uuu{\5\m}{\rangle}{0}
\zzz{-0.05cm}
\www{
\uuu{\5\tt}{\z{[}}{2.8}\m
\uuu{\2\sf}{\z{R}}{3.5}\m
\uuu{\5\tt}{\z{]}}{2.8}
}{4}
\4=\6
0
\zzz{1cm}
\z{
for 
\zzz{0.2cm}
$\zzzzzz
\uuu{\m}{\alpha}{0}
\mn\2(\n\4\rho\n\3,\nnn\1\sigma\nn\5\z{\bf ;}\n\6\xi\n\3,\nnn\1\eta\3\nnnn)
\4=\m-\mnn\m1
$}
\z{\7.}
\ee

\SEC{Lattice Versions for the Energy\1-Momen\nnn-\zyx tum Tensor}
\label{Sec_Lattice_EM_Tensors}

(3.9) up to (3.15) and (3.18) up to (3.21) have been formulated for 
\z{\12\2-\1by\2-\12\1} matrices whereas all of the other formulae 
in \sn{3} are destined to the gen\nnn-\xyz eral case of 
\z{\1N\1-\1by\2-\1N\1} matrices 
to be returned to now. Using the hitherto in\nnn-\xyz troduced auxiliary 
quantities, it will become convenient to compare various versions 
for the symmetric energy\1-\1momentum tensor on the lattice. The 
con\nnn-\xyz crete realization of the ideas (2.5) and (2.7) in \sn{2} 
is performed by starting with the continuum version of the 
energy\1-\1momentum tensor and by substi\nnn-\xyz

\zzzz{-1.cm}\np
tuting 
the continuum field strength tensor \vw{
{\z{\x F}}_{\zz{}{\zzz{-0.12cm}\z{\yyy\zzzzz$\mu\zzz{0.02cm}\nu$}}}
} there 
by the refinements\xyz

\zzzz{-1.cm}\np
\vw{\zzzzzz
\zzz{-0.2cm}
\begin{array}{l}
\zzz{0.4cm}\5
\z{\yy\1(\2$l$\2)}
\zzzz{-0.26cm}\\
{\zz{\z{$\4\cal F\mmmm$}}{\z{\yy$\sim$}}}
\zz{}{\mmmmm\z{\yy\m$\mu\1\nu$}}
\zzzz{-0.cm}
\vvy
\zzz{0.4mm}\uuu{\5}{\prime}{3.66}\n
} of \vw{\zzzzzz
\zzz{-0.2cm}
\begin{array}{l}
\zzz{0.4cm}\5
\z{\yy\1(\2$l$\2)}
\zzzz{-0.26cm}\\
{\zz{\z{$\4\cal F\mmmm$}}{\z{\yy$\sim$}}}
\zz{}{\mmmmm\z{\yy\m$\mu\1\nu$}}
\zzzz{-0.cm}\vvy
}, wich have been defined 
by a concatenation of both (3.1)\xyz

\zzzz{-0.86cm}\np
and (3.2) with (3.16) and a 
contingent application of (3.17). The actual ap\1\nn-\xyz

\zzzz{-0.94cm}\np
plication 
of (3.17) shall be marked by the label \z{$\4
\uuu{\6\bf}{\z{'}}{-1}\zzz{0.08cm}
\uuu{\bf}{\z{HYBRID}}{0}\zzz{0.05cm}
\uuu{\6\bf}{\z{'}}{-1}
\4$} and 
its re\2-\xyz

\zzzz{-1.cm}\np
spective omission by the 
counterassignment \z{$\4
\uuu{\6\bf}{\z{'}}{-1}\zzz{0.06cm}
\uuu{\bf}{\z{PURE}}{0}\zzz{0.066cm}
\uuu{\6\bf}{\z{'}}{-1}
\4$\m.}

A well\1-\nnn\0known lattice model for the energy\1-\1momentum 
tensor~\cite{Caracciolo:pt,Caracciolo:1991cp} can be in this mode
described by

\zzzz{-0.6cm}
\wx
{\z{\x$\Theta$}}
^{\zz{\z{\yyy\sf c\nnn\0a\nnn\0r\nnn\0a}}{}}
_{\zz{}{\mu\2\nu}}\3
\3=\3
\uuu{\7}{(}{-1.66}
\4
\vvx\z{\xx$\w{1}{2}$}\vvy\6
\delta\mnn\m\vvv{\mu\2\nu}{-2}\6
\delta\2\nnnn\vvv{\rho\3\xi}{5}\6
\delta\2\nnnn\vvv{\sigma\2\eta}{5}
\4+\4
2\6
\delta\mnn\0\vvv{\mu}{-2}\mmm\m\vvv{\rho}{5}\6
\delta\2\nnnn\vvv{\sigma\2\nn\xi}{5}\6
\delta\mnn\1\vvv{\nu}{-2}\mm\mnnn\m\vvv{\eta}{5}
\5
\uuu{\7}{)}{-1}
\6
\uuu{\5}{\cdot}{0}
\wy

\zzzz{-1.cm}
\be
\uuu{\5}{\cdot}{0}
\7
\trace
\3
\z{$\zzzzzz
\uuu{\7}{(}{-1}
\4
\zzz{-0.2cm}\begin{array}{l}\zzz{0.4cm}\5\z{\yy\1(\11\1)}
\zzzz{-0.26cm}\\{\zz{\z{$\4\cal F\mmmm$}}{\z{\yy$\sim$}}}
\zz{}{\mmmmm\z{\yy\m\1$\rho\1\sigma$\m}}\zzzz{-0.cm}\vy\zzz{-0.2cm}
\zzz{0.6mm}\uuu{\5}{\prime}{3.66}
\6
\zzz{-0.2cm}\begin{array}{l}\zzz{0.4cm}\5\z{\yy\1(\11\1)}
\zzzz{-0.26cm}\\{\zz{\z{$\4\cal F\mmmm$}}{\z{\yy$\sim$}}}
\zz{}{\mmmmm\z{\yy\m\2$\xi\1\eta$\mm}}\zzzz{-0.cm}\vy\zzz{-0.2cm}
\zzz{0.6mm}\uuu{\5}{\prime}{3.66}
\6
\uuu{\7}{)}{-1}
$}
\9
\stroke
\mmm
\uuu{\1\bf}{\z{HYBRID}}{7}
\ee

\zzzz{-0.4cm}
\np 
so 
that \vw{\zzzzzz
{\z{\x$\Theta$}}
^{\zz{\z{\yyy\sf c\nnn\0a\nnn\0r\nnn\0a\1\bf PUR\n\0E}}{}}
_{\mm\n\zz{}{\mu\1\nnnn\nu}}
\1
} is 
different 
from \z{\vww{\zzzzzz
{\z{\x$\Theta$}}
^{\zz{\z{\yyy\sf c\nnn\0a\nnn\0r\nnn\0a}}{}} 
_{\mm\n\zz{}{\mu\1\nnnn\nu}}
\1
\equiv
\1
{\z{\x$\Theta$}}
^{\zz{\z{\yyy\sf c\nnn\0a\nnn\0r\nnn\0a\1\bf HYB\mn\0RID}}{}}
_{\mm\n\zz{}{\mu\1\nnnn\nu}}
\1
}\m.} Switching 
from\xyz

\zzzz{-0.88cm}\np 
(2.5) to (2.7), we obtain the totally new lattice model

\yzy 

\zzzz{-0.95cm}
\be
{\z{\x$\Theta$}}
^{\zz{\z{\yyy\sf o\zzz{0.14mm}w\zzz{0.19mm}n}}{}}
_{\zz{}{\mu\2\nu}}\3
\1\defin\2
\uuu{\7}{(}{-1.66}
\4
\vvx\z{\xx$\w{1}{2}$}\vvy\6
\delta\mnn\m\vvv{\mu\2\nu}{-2}\6
\delta\2\nnnn\vvv{\rho\3\xi}{5}\6
\delta\2\nnnn\vvv{\sigma\2\eta}{5}
\4+\4
2\6
\delta\mnn\0\vvv{\mu}{-2}\mmm\m\vvv{\rho}{5}\6
\delta\2\nnnn\vvv{\sigma\2\nn\xi}{5}\6
\delta\mnn\1\vvv{\nu}{-2}\mm\mnnn\m\vvv{\eta}{5}
\5
\uuu{\7}{)}{-1}
\7
\z{\sl trace}
\3
\z{$\zzzzzz
\uuu{\7}{(}{-1}
\4
\zzz{-0.2cm}\begin{array}{l}\zzz{0.4cm}\5\z{\yy\1(\12\1)}
\zzzz{-0.26cm}\\{\zz{\z{$\4\cal F\mmmm$}}{\z{\yy$\sim$}}}
\zz{}{\mmmmm\z{\yy\m\1$\rho\1\sigma$\m}}\zzzz{-0.cm}\vy\zzz{-0.2cm}
\zzz{0.6mm}\uuu{\5}{\prime}{3.66}
\6
\zzz{-0.2cm}\begin{array}{l}\zzz{0.4cm}\5\z{\yy\1(\12\1)}
\zzzz{-0.26cm}\\{\zz{\z{$\4\cal F\mmmm$}}{\z{\yy$\sim$}}}
\zz{}{\mmmmm\z{\yy\m\2$\xi\1\eta$\mm}}\zzzz{-0.cm}\vy\zzz{-0.2cm}
\zzz{0.6mm}\uuu{\5}{\prime}{3.66}
\6
\uuu{\7}{)}{-1}
$}
\z{\7,}
\ee

\zzzz{-0.6cm}
\np 
leaving undecided initially 
whether \vw{\zzzzzz
{\z{\x$\Theta$}}
^{\zz{\z{\yyy\sf o\zzz{0.14mm}w\zzz{0.19mm}n\n\1\bf PUR\n\0E}}{}}
_{\mm\n\zz{}{\mu\1\nnnn\nu}}
\1
} or \vw{\zzzzzz
{\z{\x$\Theta$}}
^{\zz{\z{\yyy\sf o\zzz{0.14mm}w\zzz{0.19mm}n\n\1\bf HYB\mn\0RID}}{}}
_{\mm\n\zz{}{\mu\1\nnnn\nu}}
\1
} has 
to be\xyz

\zzzz{-0.95cm}\np 
preferred. A third model with 
intermediate properties~\cite{Karsch:1986cq} uses 
the hetero\1\nnn-\xyz geneous construction principle

\zzzz{-0.56cm}
\alphaon
\wx
\zzz{-0.22cm}      
{\z{\x$\Theta$}}
^{\zz{\z{\yyy\sf k\nnn\0a\nnn\0r\nnn\0s\nnn\0c\nnn\0h}}{}}
_{\zz{}{\mu\2\nu}}
\5=\3
-\3
\w{\www{2}{-1}
}{\1
\vvx\zzzz{-0.64cm}\\{\kappa}\vvv{\nnnn\12}{4}\vvy 
\2}
\8
\z{\sl trace}
\4
\uuu{\6}{(}{-1}
\1
-
\4
\sum_{\lambda}
\zzz{-0.71cm}
\vvv{\lambda\1\neq\1\mu}{-15}
\zzz{0.1cm}
\z{U}
\uuu{\2}{\mm\mnn\mu\2\lambda}{-3}
\8
+
\6
\sum_{\sigma\2,\4\mnn\lambda}
\zzz{-0.76cm}
\vvv{\sigma\0,\mnn\lambda\1\neq\1\mu
\6\z{\bf,}\7 
\sigma
\2\uuu{\0}{>}{0.5}\2\nn
\lambda}{-15}
\zzz{0.05cm}
\z{U}
\uuu{\2}{\mm\nn\sigma\nnnn\lambda}{-3}
\9
\uuu{\6}{)}{-1}
\6
+
\6
\vvv{\z{\x\zzzzz$\cal O$}}{-1}\2
\uuu{\m}{(}{0.6} 
\4
\kappa\1
\vvv{2}{4}
\6\mnnn
\uuu{\m}{)}{0.6}
\wy

\zzzz{-1cm}
\be
\zzz{9.198cm}
\z{\x\tt for}
\zzz{0.5cm}
\mu\2=\3\nu 
\ee

\zzzz{-0.95cm}
\np 
and

\zzzz{-0.5cm}
\be
\0
\zzz{-1.186cm}
{\z{\x$\Theta$}}
^{\zz{\z{\yyy\sf k\nnn\0a\nnn\0r\nnn\0s\nnn\0c\nnn\0h}}{}}
_{\zz{}{\mu\2\nu}}
\4=\2
-\12
\7
\delta\2\nnnn\vvv{\sigma\1\mnn\lambda}{5}
\7
\z{\sl trace}
\3
\z{$\zzzzzz
\uuu{\7}{(}{-1}
\4
\zzz{-0.2cm}\begin{array}{l}\zzz{0.4cm}\5\z{\yy\1(\11\1)}
\zzzz{-0.26cm}\\{\zz{\z{$\4\cal F\mmmm$}}{\z{\yy$\sim$}}}
\zz{}{\mmmmm\z{\yy\m\1$\mu\1\nnn\sigma$\m}}\zzzz{-0.cm}\vy\zzz{-0.2cm}
\zzz{0.6mm}\uuu{\5}{\prime}{3.66}
\6
\zzz{-0.2cm}\begin{array}{l}\zzz{0.4cm}\5\z{\yy\1(\11\1)}
\zzzz{-0.26cm}\\{\zz{\z{$\4\cal F\mmmm$}}{\z{\yy$\sim$}}}
\zz{}{\mmmmm\z{\yy\m\2$\nu\nn\lambda$\mm}}\zzzz{-0.cm}\vy\zzz{-0.2cm}
\zzz{0.6mm}\uuu{\5}{\prime}{3.66}
\6
\uuu{\7}{)}{-1}
$}
\zzz{1.1cm}
\z{\x\tt for}
\zzz{0.5cm}
\mu\2\neq\3\nu
\z{\7.\zzz{-0.28cm}\0}
\zzz{-0.5cm}
\ee
\alphaoff

\SEC{Monte Carlo Results}
\label{Sec_MC_Results}

The Euclidean expectation values for the components of the presented 
lattice models for the symmetric energy\1-\1momentum 
tensor \vw{{\Theta}_{\mm\zz{}{\mu\1\nu}}} have been measured\xyz

\zzzz{-0.97cm}\np
on \z{a 
\z{$\4{\m1\mnn0}\mn\3\uuu{\4\m}{*\nnn*}{0.2}\mn\44\2\nnn$} 
lattice.} For this purpose, a heat bath   
\z{\n\0SU(2)\n} Monte Carlo simu\1-\xyz lation~\cite{Creutz:zw} 
has been performed, measuring once 
every 50 sweepings after a cold start with a transient state  
of \z{1\m0\2\nnn000} lattice updatings. Basically 
the average concerning all of the lattice sites and every 
performance of a measurement of the quantity of reference there 
until a preliminarily most recent 
lattice sweeping \vw{X} is defined as the Monte Carlo output 
read at sweep \vw{X}. The 
Figures\fs5.1\fs\0and\fs5.2 
compare the ground state expectation values for a 
rep\nn\1-\xyz resentative 
\z{off\nnn\1-}\1diagonal component 
and for a likewise vicarious 
(\1it could be 
\vw{{\Theta}_{\mm\zz{}{4\14}}} 
as well, by reason of 
\z{\n\0O\nn(4)\n} invariance 
in the regarded system\1) 
diagonal\xyz

\zzzz{-0.97cm}\np 
component 
of \vw{{\Theta}_{\mm\zz{}{\mu\1\nu}}}, respectively, with \vww{
\hat{\beta}
\1=\1    
2\2\z{N} 
\1/\1    
g_{\m\zz{}{\circ}}^{\32}
} (the 
hat of \vw{\hat{\beta}} is dropped\xyz

\zzzz{-0.95cm}\np
when paraphrased by the Latin 
capitals "BETA" in the plot data captions inside the 
Figs.\fs5.1\fs\0and\fs5.2 
themselves) for \vww{
\z{N}
=\1
2
} as 
close as possible to the domain of the 
\z{\n\0SU(2)\n} scaling window.
\begin{figure} 
\0\zzzz{-0.7cm}
\begin{center}
\Fig{1}{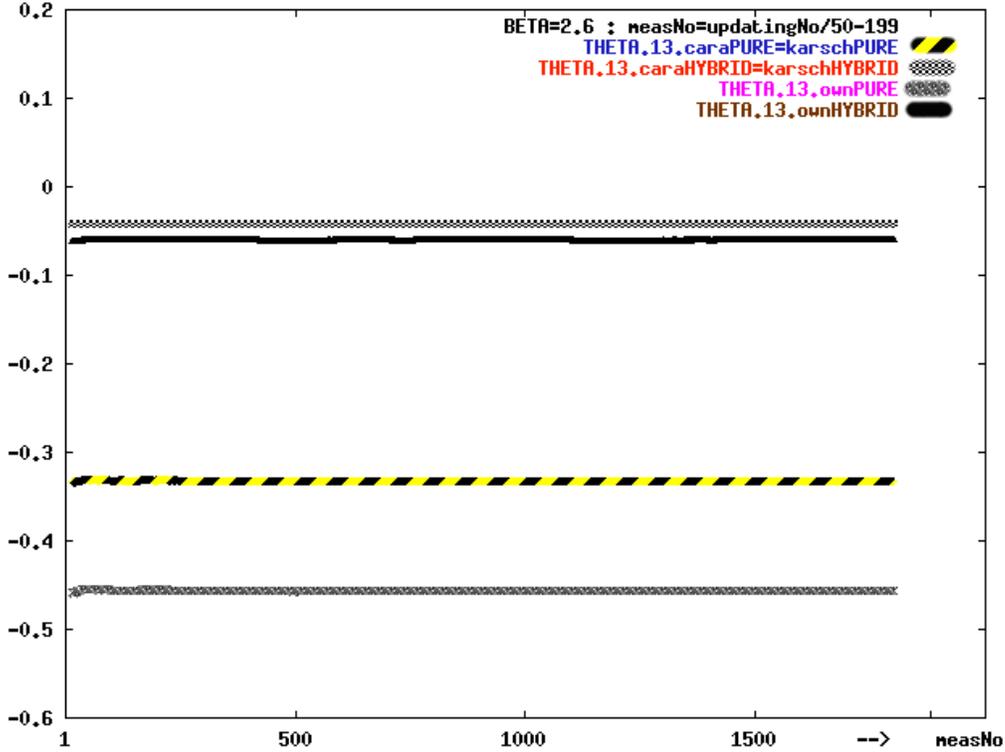}{270}
\end{center}
\caption[10pt]{   
  The Euclidean ground state expectation values for an arbitrarily\xyz  
  chosen \z{off\nnn\1-\1diagonal} component of the symmetric 
  energy\1-\1momentum tensor\xyz are plotted as a function of the number 
  of relevant lattice updatings executed so far. The first 
  \z{1\m0\2\nnn000}~(\29999\2) lattice updatings are discarded and 
  ignored when counting the total number of measurements and the 
  same disregard\1-\xyz ing is for the sakes of pseudo\nnn\1-\1decorrelation  
  bound to happen to every 49 of respective 50 subsequent 
  iterations each. There are effectively four diverse\xyz graphs,   
  resulting from six different constructions for the whole 
  \z{energy\1-\1mo\1\nnn-}\xyz mentum tensor on the lattice which are 
  discussed in the main part of the text.
}
\label{Fig_5.1_Theta_13_(measNo,modelType,beta=2.6)}
\end{figure}
\begin{figure} 
\0\zzzz{-3.7cm}
\begin{center}
\parbox{\textwidth}{
\bp{137}{10}{1}\put(-0.04,0.75){
\Fig{1}{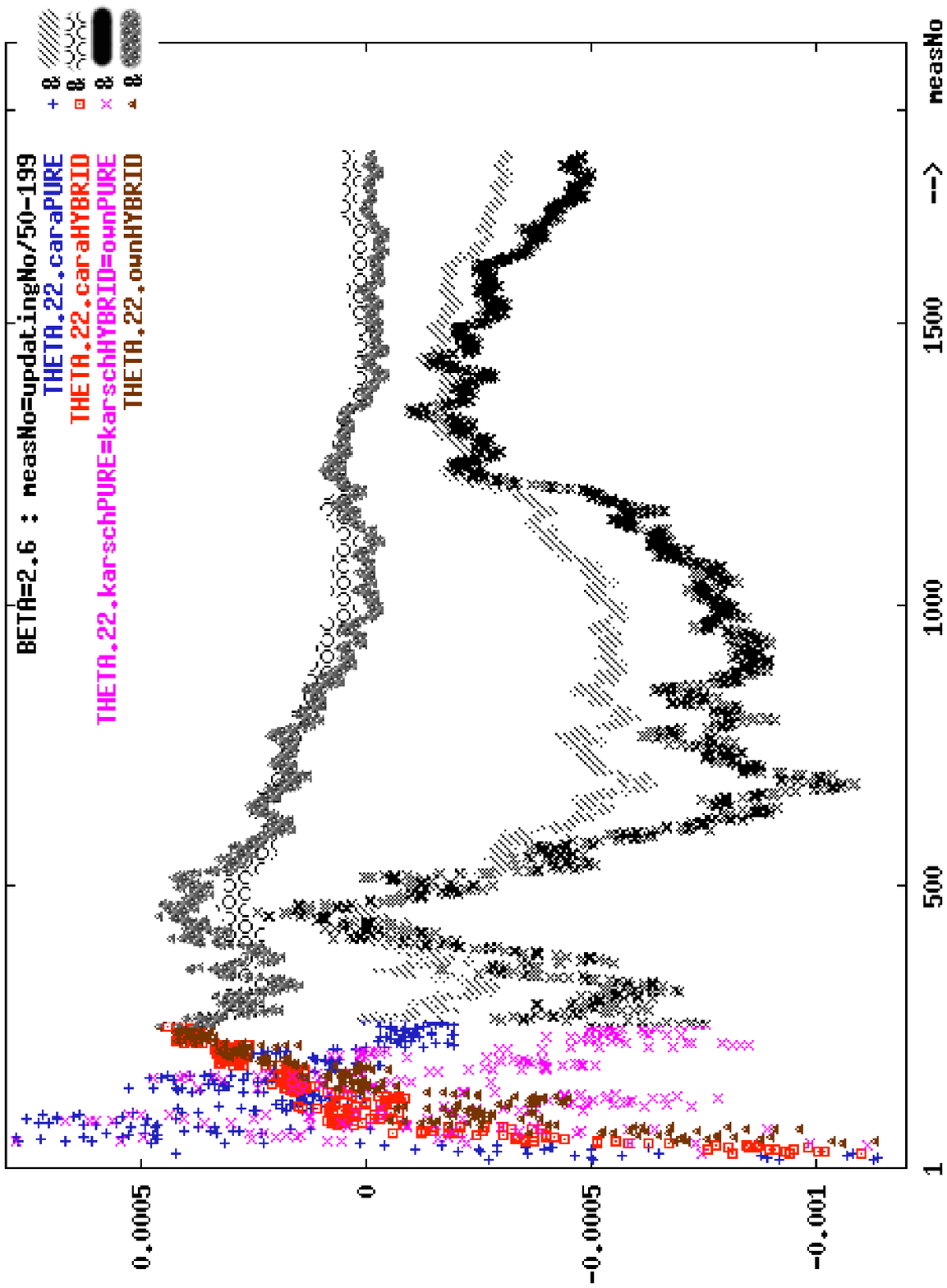}{270}
}\ep
}
\end{center}
\caption[10pt]{   
  The same scenario as in Fig.\fs5.1\1, but for an arbitrarily selected 
  diagonal component of the energy\1-\1momentum tensor this time. Again, 
  six lattice models merge into four effective graphs, but in a manner 
  that is unlike that one concerning Fig.\fs5.1\1. Statistical 
  fluctuations, referring to the applied Monte Carlo algorithm, are 
  better visible than in Fig.\fs5.1 because the scale for the axis of 
  ordinates of the displayed plot is much smaller here.  
}
\label{Fig_5.2_Theta_22_(measNo,modelType,beta=2.6)}
\end{figure}

In four dimensions, the 
investigated \vw{\zzzzzz
\z{\sl\nnn\0a\mn\0b\nnn\0s}\4 
\m\uuu{\5\m}{\langle}{0}  
\5
\mm{\Theta}_{\n\zz{}{2\2\mnnn2}}\mm
\1
\5
\uuu{\5\m}{\rangle}{0}\m
} has 
to thermalize to\nnn\1-\xyz

\zzzz{-0.5cm}
\zzzz{-1.03cm}\np 
wards zero in each model 
presented above because  
\z{\n\0O\nn(4)\n} invariance 
effec\1-\xyz tively supplies equal numbers of terms and analogous 
subtractive counter\1-\xyz terms then. On the other hand, a 
model\1-\nnn\0dependent lattice simulation 
of\xyz

\zzzz{-1.02cm}\np 
\m\vw{\zzzzzz
\z{\sl\nnn\0a\mn\0b\nnn\0s}\4 
\m\uuu{\5\m}{\langle}{0}  
\5
\mm{\Theta}_{\n\zz{}{\nn1\1\mn3}}\mm
\1
\5
\uuu{\5\m}{\rangle}{0}\m
} is the better the more 
its limiting value lies in vicinity to 
the\xyz

\zzzz{-1.02cm}\np
corresponding vanishing Euclidean vacuum expectation value 
for \vw{\zzzzzz 
{\Theta}_{\n\zz{}{\nn1\1\mn3}}
} in\xyz

\zzzz{-1.03cm}\np 
continuum physics. Since the actually used lattice itself is fairly 
too coarse\xyz

\yzy 

\np 
\z{w\mm.\1r\m.\1t.} its spacing and much too small 
regarding the extension of one period\1-\xyz icity segment
Figs.\fs5.1\fs\0and\fs5.2 
reveal that the most convincing results 
for\xyz

\zzzz{-0.98cm}\np
\m\nnn\vw{\zzzzzz
\m\uuu{\5\m}{\langle}{0}  
\5
\mm{\Theta}_{\n\zz{}{\nn1\1\mn3}}\mm
\1
\5
\uuu{\5\m}{\rangle}{0}\m
} and \vw{\zzzzzz
\m\uuu{\5\m}{\langle}{0}  
\5
\mm{\Theta}_{\n\zz{}{2\2\mnnn2}}\mm
\1
\5
\uuu{\5\m}{\rangle}{0}\m
} are obtained if the underlying model is chosen to be\xyz

\zzzz{-0.95cm}\np
of the 
Caracciolo~\cite{Caracciolo:pt,Caracciolo:1991cp} 
type \z{$\4
\uuu{\6\bf}{\z{'}}{-1}\zzz{0.04cm}
\uuu{\2\m}{\cal CARA}{0}\zzz{0.07cm}
\uuu{\6\bf}{\z{'}}{-1}
\4$} \z{(\1cf\1.~(4.1)\1)}
\z{and\nnn/\mnnn\0or} 
the 
\z{\14\2-\nnnn\0plaquettes}\xyz

\zzzz{-1cm}\np 
hybridization 
procedure \z{$\4
\uuu{\6\bf}{\z{'}}{-1}\zzz{0.08cm}
\uuu{\bf}{\z{HYBRID}}{0}\zzz{0.05cm}
\uuu{\6\bf}{\z{'}}{-1}
\4$} (3.17) 
is activated. By construction, the expectation values 
of \vw{\zzzzzz 
{\Theta}_{\n\zz{}{2\2\mnnn2}}
} are 
equal for the 
models\zyx

\zzzz{-0.92cm}\np 
\z{$\4
\uuu{\6\bf}{\z{'}}{-1}\zzz{0.04cm}
\uuu{\2\m}{\cal KARSCH}{0}\zzz{0.08cm}
\uuu{\bf}{\z{PURE}}{0}\zzz{0.05cm}
\uuu{\6\bf}{\z{'}}{-1}
\4$}, \z{$\4
\uuu{\6\bf}{\z{'}}{-1}\zzz{0.04cm}
\uuu{\2\m}{\cal KARSCH}{0}\zzz{0.08cm}
\uuu{\bf}{\z{HYBRID}}{0}\zzz{0.05cm}
\uuu{\6\bf}{\z{'}}{-1}
\4$}, and \z{$\4
\uuu{\6\bf}{\z{'}}{-1}\zzz{0.04cm}
\uuu{\2\m}{\cal OWN}{0}\zzz{0.08cm}
\uuu{\bf}{\z{PURE}}{0}\zzz{0.05cm}
\uuu{\6\bf}{\z{'}}{-1}
\4$}, im\nnn-\xyz

\zzzz{-0.88cm}\np 
plying 
that \vw{\zzzzzz
\m\uuu{\5\m}{\langle}{0}  
\5
{\z{\x$\Theta$}}
^{\zz{\z{\yyy\sf o\zzz{0.14mm}w\zzz{0.19mm}n\n\1\bf HYB\mn\0RID}}{}}
_{\nn\zz{}{2\2\mnn2}}
\1
\5
\uuu{\5\m}{\rangle}{0}\m
} can 
thermalize better 
than \vw{\zzzzzz
\m\uuu{\5\m}{\langle}{0}  
\5
{\z{\x$\Theta$}}
^{\zz{\z{\yyy\sf k\nnn\0a\nnn\0r\nnn\0s\nnn\0c\nnn\0h\1\bf HYB\mn\0RID}}{}}
_{\nn\zz{}{2\2\mnn2}}
\1
\5
\uuu{\5\m}{\rangle}{0}\m
}.

\zzzz{0.33cm}
To the 
contrary, \vw{\zzzzzz
{\z{\x$\Theta$}}
^{\zz{\z{\yyy\sf k\nnn\0a\nnn\0r\nnn\0s\nnn\0c\nnn\0h\1\bf PUR\n\0E}}{}}
_{\nn\zz{}{1\13}}
\1=\1
{\z{\x$\Theta$}}
^{\zz{\z{\yyy\sf c\nnn\0a\nnn\0r\nnn\0a\1\bf PUR\n\0E}}{}}
_{\nn\zz{}{1\13}}
\1
} and\zyx

\zzzz{-0.8cm}\np 
\vw{\zzzzzz
{\z{\x$\Theta$}}
^{\zz{\z{\yyy\sf k\nnn\0a\nnn\0r\nnn\0s\nnn\0c\nnn\0h\1\bf HYB\mn\0RID}}{}}
_{\nn\zz{}{1\13}}
\1=\1
{\z{\x$\Theta$}}
^{\zz{\z{\yyy\sf c\nnn\0a\nnn\0r\nnn\0a\1\bf HYB\mn\0RID}}{}}
_{\nn\zz{}{1\13}}
\1
} are 
direct consequences of (4.1) and (4.3\2b).\xyz

\zzzz{-0.82cm}\np  
Thus \vw{\zzzzzz
\z{\sl\nnn\0a\mn\0b\nnn\0s}\4 
\m\uuu{\5\m}{\langle}{0}  
\5
{\z{\x$\Theta$}}
^{\zz{\z{\yyy\sf k\nnn\0a\nnn\0r\nnn\0s\nnn\0c\nnn\0h\1\bf HYB\mn\0RID}}{}}
_{\nn\zz{}{1\13}}
\1
\5
\uuu{\5\m}{\rangle}{0}\m
} is 
closer to zero 
than \vw{\zzzzzz
\z{\sl\nnn\0a\mn\0b\nnn\0s}\4 
\m\uuu{\5\m}{\langle}{0}  
\5
{\z{\x$\Theta$}}
^{\zz{\z{\yyy\sf o\zzz{0.14mm}w\zzz{0.19mm}n\n\1\bf HYB\mn\0RID}}{}}
_{\nn\zz{}{1\13}}
\1
\5
\uuu{\5\m}{\rangle}{0}\m
}.\xyz

\zzzz{-0.98cm}\np 
Nevertheless, 
there is no real need 
for \vw{\zzzzzz
\m\uuu{\5\m}{\langle}{0}  
\5
\mm{\Theta}_{\n\zz{}{\nn1\1\mn3}}\mm
\1
\5
\uuu{\5\m}{\rangle}{0}\m
} to vanish on the lattice because\xyz

\zzzz{-1.01cm}\np
Euclidean expectation values 
of the symmetric 
energy\1-\1momentum tensor in\nnn-\xyz deed do not have negative 
parity\1--\1or spacetime 
\z{\four\nnnn\0parity,} 
respectively. It is\xyz

\zzzz{-1.02cm}\np 
intuitively evident that the 
model \z{$\4
\uuu{\6\bf}{\z{'}}{-1}\zzz{0.04cm}
\uuu{\2\m}{\cal OWN}{0}\1\zzz{0.07cm}
\uuu{\6\bf}{\z{'}}{-1}
\4$} (4.2) should 
at least exhibit\xyz

\zzzz{-1.02cm}\np
slightly deteriorated numerical 
convergence properties relative  
to \z{$\4
\uuu{\6\bf}{\z{'}}{-1}\zzz{0.04cm}
\uuu{\2\m}{\cal CARA}{0}\zzz{0.07cm}
\uuu{\6\bf}{\z{'}}{-1}
\4$} (4.1) in 
the lattice simulation, due to the sensitively more 
sophisticated alge\2-\xyz braic realization of the 
half\nnn\1-\nnn\0angle concept (3.4) by (3.2). This is in point 
of fact what is observed. Monte Carlo simulation improvements by 
\z{\14\2-\nnnn\0plaquettes} averaging 
are throughout not inhibited if the model of reference  
possesses a uniform construction principle for each component of 
the energy\1-\1momentum tensor, like (4.1) and (4.2).

\SEC{Discussion}
\label{Sec_Discussion}

The mathematical contents of an arbitrary Yang\1-Mills theory (we do 
not care about principles of symmetry breaking here) is just 
the nonabelian general\nnn-\xyz ization of structures 
that we classically know 
in the form of electromagnetism. Ref.~\cite{Holk:2003qa} demonstrates 
that these structures are pure \z{"\m\0whirl\2"} structures 
bas\n\1-\xyz ing 
upon appropiately generalized cross products and curl operations so 
that we tend to infer from this insight solely that these structures 
are not suited to \z{"\m\mnn\0arrive\2"} at the symmetric 
energy\1-\1momentum tensor by a genuine access, like differential 
forms, e.\1g.\1. The same phenomenon is reflected by the 
circum\nnn-\xyz stance 
that the only purely special relativistic Noether current 
representation

\zzzz{-0.66cm}
\be
{\Theta}\1\nnn\uuu{\1}{\mu\1\nu}{4}
\2=\2
g^{\nnnn\mu\1\nu}\3
{\cal L}
\2-\2
g^{\nnnn\mu\1\lambda}\4
\w{\partial\1{\cal L}}
{
\partial\3
{\partial}_{\mnn\zz{}{\nu}}\3
{\varphi}^{\1a}\2
}\5
{\partial}_{\m\zz{}{\lambda}}\2
{\varphi}^{\1a}
+\1
{\partial}_{\m\zz{}{\lambda}}\1
{\mit\Sigma}^{\1\mu\1\nu\1\m\lambda}
\ee

\zzzz{-0.44cm}\np 
of the symmetric energy\1-\1momentum tensor does not supply 
\z{\2D\2-\1dimensional}\xyz

\yzy 

\zzzz{-1cm}\np
information for the structure of the trace anomaly 
because \vww{
{\mit\Sigma}^{\1\mu\1\nu\1\m\lambda}
\1=\n-\nnn
{\mit\Sigma}^{\1\mu\1\lambda\1\nn\nu}\1
}\xyz

\zzzz{-0.98cm}\np 
has to be fitted "by hand\2" for 
implying \z{\vww{\zzzzzz 
{\Theta}\1\nnn\uuu{\1}{\mu\1\nu}{4}
\mnn\1=\1
{\Theta}\nnnnn\uuu{\1}{\nu\mu}{4}
}.} \vww{
{\mit\Sigma}^{\1\mu\1\nu\1\m\lambda}
\1=\2
0
} would 
mere\2-\xyz

\zzzz{-1.02cm}\np
ly 
give the canonical energy\1-\1momentum tensor, referring to 
the investigated spacetime translation invariance. 

This difficulty can be algebraically overcome by structures with 
more re\2-\xyz fined combinations of the 
\z{Levi\nnn\1-\nnn\0Civita} tensor. The continuation of 
\z{self\nnn\1-\nnn\0simi\1-}\xyz larity 
arguments to the domain of 
general relativity in Ref.~\cite{Holk:2003qa} is associated with 
the gauge group \z{\n\0SO\nn(4)\n}, whose structure constants can 
be in this way\xyz

\zzzz{-1.02cm}\np
described by elementary spacetime tensors 
exclusively 
if \z{$\5\zzzzzz 
\varphi\2(\2\z{A}\1)\defin\1\z{a}
\5$} and\xyz

\zzzz{-1.02cm}\np
\z{$\5\zzzzzz 
\psi\2(\2\z{A}\1)\defin\1\z{b}
\5$} are 
the inverse index\1-index functions relative to

\zzzz{-1.1cm}
\be
\z{A\2(\1a\4,\4b\2)}
\2\defin\2 
\z{\sl ma\1x}\2
\z{\bf(}\2
\z{a}\2+2\4\z{b}-5
\4\z{\bf,}\4
1\2
\z{\bf)}
\9\6\forall\9\1 
\z{a}\3,\1\z{b}\6\in\{\21\1,\1.\1.\1.\2,\14\2\} 
\zzz{0.2cm}\z{with}\zzz{0.2cm}\z{b}>\z{a}\5,
\ee

\zzzz{-0.5cm}\np
obeying \z{$\5\zzzzzz 
\varphi\2(\2\z{A}\1)\1<\2\psi\2(\1\z{A}\1)
\5$}:

\zzzz{-0.66cm}
\wx
\z{\x$\zzzzzz
f_{\zz{}{\z{\yy A\nnnn\0B\nnnn\0C}}}
\zzz{-0.7cm}
\uuu{\1}{\z{\tt S\1O\m(\n\m4\n)}}{6}
\2\equiv\mmm 
\1-\22\1\5i\7
\trace\4
\z{\bf(}\4
\minus{  
{\hat{\tau}}_{\!\zz{}{\z{\yy A}}}
}{
{\hat{\tau}}_{\!\zz{}{\z{\1\yy B}}}
} 
\6{\hat{\tau}}_{\!\zz{}{\z{\1\yy C}}}
\4\z{\bf)}
\1=\3
\uuu{\4}{\zzzzzz\w{\www{1}{-1.4}}{2}}{1.15}
\4
\4{\delta}^{\4\zz{\z{\y a\3g}}{}}
\4{\delta}^{\4\zz{\z{\y c\3e}}{}}
\4{\delta}^{\4\zz{\z{\y d\3f}}{}}
\6\ast
$}
\wy

\zzzz{-1.3cm}
\wx
\z{\x$\zzzzzz
\ast\3
\sum_{\5b\1=\11\5}^{4}
{\z{\xx $\varepsilon$}}_{\!\!\!\zz{}{\z{
\y $\varphi\3$({\yy A})$\3\psi\3$({\yy A})\5a\4b}}}
\6{\z{\xx$\varepsilon$}}_{\!\!\!\zz{}{\z{
\y $\varphi\3$({\yy B})$\3\psi\3$({\yy B})\5c\4b}}}
\6{\z{\xx$\varepsilon$}}_{\!\!\!\zz{}{\z{
\y $\varphi\3$({\yy C})$\3\psi\3$({\yy C})\5d\4b}}}
\6{\z{\xx$\varepsilon$}}_{\!\!\!\zz{}{\z{
\y\5e\5f\5g\5b}}}
$}
\wy

\zzzz{-0.7cm}
\be
\z{\x$\zzzzzz
\forall\3\;\z{\y A\3,\3B\3,\3C}\;\in\;\{\21\1,\1.\1.\1.\2,\16\2\}\4.
$}
\ee

\zzzz{-0.4cm}
\np
The related 
\z{$(\nn\0i\mnn\3\z{\sl c}\4t\nnn)\1$-Euclidean} 
Riemann tensor \vw{
\z{R}\2\vvv{\alpha\2\beta\2\n\gamma\2\nn\delta\1}{-2}
} is, again, in contact\xyz

\zzzz{-1.cm}\np 
with a further type of such an 
ingeniously concatenated 
combination of \z{Levi\1\nnn-\nnn\0Civita} tensors 
if exterior calculus is utilized for its remodeling into the 
Einstein tensor \vw{
\z{G}\vvv{\mu\1\n\nu\1}{-2}
}, being proportional to
the symmetric energy\1-\1momentum\xyz

\zzzz{-1.04cm}\np
tensor 
of general relativity (\1the 
\vw{\Theta\vvv{\m\m\mu\1\nu}{-1.7}\zzz{-0.31cm}\vvv{(1)}{4.2}} 
of \sn{2}\2) after evaluation of the Einstein\xyz

\zzzz{-1.02cm}\np
field equations\1:

\zzzz{-0.75cm}
\wx
\zzz{-3.75cm}
{\z{d\1x}}^{\zz{\z{\yyy$\3\alpha$}}{}}
\4
\z{\yyy$\wedge$}
\zzz{-0.25cm}
\vx
\zzz{0.16cm}\z{\yyy$\wedge$}\zzz{0.1cm}\zzzz{-0.4 cm}\\
\zzz{0.66cm}\z{\yy$\beta$}\zzzz{-0.35cm}\\
{\z{e}}\zzzz{0.35cm}
\vy
\ub(\mmm-\mm\z{\x$\w{1}{4}$}\4\ub)
\6
{\z{\x$\varepsilon$}}_{\m\zz{}{\alpha\2\beta\2\n\gamma\2\nn\delta}}
\9
{\z{R}}
_{\zzz{0.46cm}\zz{}{\z{\yyy$\m\mu\2\nu$}}}
^{\zz{\z{\yyy $\3\gamma\2\delta$}}{}}
\9\2
{\z{d\1x}}^{\zz{\z{\yyy$\3\mu$}}{}}
\2
\z{\yyy$\wedge$}
\6
{\z{d\1x}}^{\zz{\z{\yyy$\3\nu$}}{}}
\4
=
\wy

\zzzz{-1.44cm}
\be
=
\zzz{-0.32cm}
\vx
\zzz{0.16cm}\z{\yyy$\wedge$}\zzz{0.1cm}\zzzz{-0.4cm}\\
\zzz{0.66cm}\z{\yy$\beta$}\zzzz{-0.35cm}\\
{\z{e}}\zzzz{0.35cm}
\vy
\ub(\mmm-\mm\z{\x$\w{1}{4}$}\4\ub)
\6
{\z{\x$\varepsilon$}}
_{\m\zz{}{\alpha\2\beta\2\n\gamma\2\nn\delta}}
\9
{\z{R}}
_{\zzz{0.46cm}\zz{}{\mbox{\yyy$\m\mu\2\nu$}}}
^{\zz{\z{\yyy$\3\gamma\2\delta$}}{}}
\9\2
{\z{\x$\varepsilon$}}^{\2\alpha\2\lambda\2\mu\2\1\nu}
\6\3
{\z{d}}^{\13}\2{\z{\yyy$\cal O$}}_{\zz{}{\m\lambda}}
\4
=
\zzz{-0.15cm}
\vx
\z{\yyy$\wedge$}\zzzz{-0.26cm}\\
\z{e}\zzzz{-0.3 cm}\\
\zzz{0.4 cm}\z{\yy$\m\sigma$}\zzzz{0.1cm}
\vy
\2
{\z{G}}^{\zz{\z{\yyy$\2\rho\1\nn\sigma$}}{}}
\4\4
{\z{d}}^{\13}\2{\z{\yyy$\cal O$}}_{\zz{}{\m\rho}}
\ee

\zzzz{-0.4cm}
In contrast to Yang\1-Mills theories, the geometrodynamic 
representation of general relativity is spanned by two sorts of 
base systems, being differential\xyz

\zzzz{-1.08cm}\np  
elements \vw{
\z{d\1x}\1\nn\vvv{\alpha}{4}
} and 
Cartan base 
vectors \vw{
\zzz{-0.52cm}\vx
\zzz{0.16cm}\z{\yyy$\wedge$}\zzz{0.1cm}\zzzz{-0.4 cm}\\
\zzz{0.66cm}\z{\yy$\mm\beta$}\zzzz{-0.35cm}\\
{\2\z{e}}\zzzz{0.266cm}
\vvy
}, both kinds of them displayed\xyz

\zzzz{-1.08cm}\np
in (6.4). This peculiarity may be 
seen in context with (6.3) and genuinely procreates symmetric tensors 
of the second rank in the framework of exterior calculus. The 
corresponding Noether current representation uses a variation of 
the Lagrangian density relative to the metric tensor and this 
procedure can be extended to Yang\1-Mills theories, acting as a 
cryptic (a posteriori invisible) general relativistic transit that 
specifies the trace anomaly of the symmetric 
energy\1-\1momentum tensor there appropiately.

\yzy 

Therefore a fundamental comprehension of the symmetric 
energy\1-\1mo\1\nnn-\xyz mentum 
tensor \vw{{\Theta}_{\mm\zz{}{\mu\1\nu}}} basically has something 
to do with general relativity.\xyz

\zzzz{-1cm}\np
The pattern (2.8) in \sn{2} is a 
continuation of this aspect, promoting (4.2) as a new lattice ansatz 
for \vw{{\Theta}_{\mm\zz{}{\mu\1\nu}}}. The average expectation value 
for the trace\xyz

\zzzz{-1.02cm}\np
anomaly of (4.2) would formally be compatible with the 
description of the trace of the energy\1-\1momentum tensor in general 
relativity referring to the extent of deviation from an 
ultrarelativistic ideal\1-\1gas state of massless gauge bosons (which 
is comparatively least unrealistic for the non\1-\1selfinteracting 
gauge group U(1)\1--\2or perhaps for the deconfined high\mn\1-temperature 
regime of QCD, as well as for low\1-temperature gluonium 
states~\cite{Engels:1981wu,Engels:1981qx,Karsch:2003zq,Kuti:1980gh}\2) 
if the isotropic special case (realizable by gases) is considered for 
position space, with the stress tensor becoming a product of pressure 
\vw{p} and the unit ma\1-\xyz trix in the comoving system (\1whereas 
its trace is of course independent of the choice of the physical 
reference system\1--\nnn\1a completely finite  
\z{$(\13\m+\m1\1)\1$-\1lattice}\xyz with 
spacings 
\z{\1"\m$a$\2\n"\m}
and \vw{
a_{\1\zz{}{\tau}}
} and 
an\1--\1inversely defined\1--\nnn\1anisotropy 
parameter \vww{
\xi\1=\1
a_{\1\zz{}{\tau}}
/a
} is 
suitable to the scenario of a field theory at finite physical 
temperature \vw{T}, where~\cite{Hosoya:1983id,Sciama:hr} 
the integration scale for 
Euclidean \z{"time\2"}, \z{\1"\m\n$y$\2\n"\m}, in the 
action runs 
from \vw{0} to \vw{1/\1T\1} 
for \z{\vww{
\z{\sl c}
=
\z{$\hbar$}
=
{\z{\xy\sl k}}_{\zz{}{\z{\yyy\sl B}}}
=
1
})\nnn:}

\zzzz{-0.6cm}
\samepage{
\wx
\zzz{0.17cm}{\White{\Theta}}
^{\zzz{0.01cm}\zzz{0.05mm}\zz{\z{\yyy$\mu$}}{}}
_{\zzz{-0.01cm}\zzz{0.05mm}\zz{\z{\yyy$\mu$}}{}}\zzz{-0.7cm}
{\White{\Theta}}^{\mm\zz{\z{\yyy\tt\2L}}{}}\zzz{-0.475cm}
{\z{\zzzzz$\bar{\z{\uuuuu$\Theta$}}$}}\zzz{0.4cm}
\defin\2
\z{\x\zzzzz$\varepsilon$}
\m-\03\3p
\4=  
\m-\1\w{\1x\3y}{\m\z{\sf V}\0
}\8\w{\2\partial\1\ln\0Z(\0 \z{\sf\1V\m} \m=\m\z{\tt const}
\0=\0x^{\13}\4y\2\propto\2x^{\13}\mm/\1T\2)\1}
{\partial\1(\zzz{0.05mm}\m x\3y\m\zzz{0.1mm})}\4=
\wy

\zzzz{-1.3cm}
\be
\zzz{8.7718cm}
=\4 
\w{\2\zzz{-0.18cm}\vx a\zzzz{-0.12cm}\vy\zzz{-0.18cm}\1}{\z{\sf V}}
\5
\z{\zzzzz$\langle$}\3
{\partial}_{\zzz{-0.01cm}\zz{}{a}}\3
{\z{\zzzzz$S$}}_{\mm\zz{}{\z{\yyy\sf G}}}\zzz{0.15mm}
\3
{\z{\zzzzz$\rangle$}}
\zzz{-0.245cm}\1\vx
\z{\yyy$\xi=\z{\yyy\sf const} 
$}\zzzz{0.6cm}
\vy
\zzz{-1.2cm}
\ee
}

\zzzz{-0.3cm}
For deriving (6.5), we have used the lattice transfer (thereby 
suppress\1-\xyz ing a ground\1-\1state energy normalization and 
integrations over constant field 
configurations\1--\1cf\1.~\cite{Engels:1981ab,Engels:1981qx}\2) of the 
statistical\1-\1physics relationship between the partition 
function \vw{Z} and the ground\1-\1state expectation value

\zzzz{-1.15cm} 
\be
\z{\zzzzz$\langle$}\4
{\partial}_{\mmmmm\zz{}{\cal A}}\7
{\z{\zzzzz$S$}}_{\mm\zz{}{\z{\yyy\sf G}}}
\3\z{\zzzzz$\rangle$}
\defin\1
\z{\zzzzz$\langle$}\3\Omega
\4\z{\zzzzz$|$}\5
{\partial}_{\mmmmm\zz{}{\cal A}}\5
{\z{\zzzzz$S$}}_{\mm\zz{}{\z{\yyy\sf G}}}
\4\z{\zzzzz$|$}\4
\Omega\3\z{\zzzzz$\rangle$}
\3=\4
\w{\zz{}{1}}{Z}\1
\int\m
{\cal D}\0
U\4
(\4
{\partial}_{\mmmmm\zz{}{\cal A}}\6
{\z{\zzzzz$S$}}_{\mm\zz{}{\z{\yyy\sf G}}}
\4)\5
{\z{\sl e}}^
{-{\z{\yyy\zzzzz$S$}}
}
\zzz{-0.2cm}\vx
\z{\yyyy\sf G}
\zzzz{0.2cm}
\vy
\0=\m-\1
{\partial}_{\mmmmm\zz{}{\cal A}}\2
\ln\0Z
\ee

\zzzz{-0.6cm}
\np 
relative to any suited quantity of reference 
\vw{{\cal A}} \z{(\1\vw{{\cal D}U\4}\mmm} is just the appropiate\xyz  
integration measure and does not directly concern the lattice quantity 
\vw{{\z{U}}_{\mm\zz{}{\z{\yy$\mu\1\nu$}}}\n}).\xyz 

\zzzz{-0.99cm}\np
Repeating 
this for \vww{{\cal A}\1=\1\xi} (at 
fixed couplings and for adjusted ground\1-\1state\xyz

\zzzz{-0.97cm}\np
energy normalization, 
v. \cite{Engels:1980ty,Engels:1981qx,Montvay:1981jj}\2) gives

\zzzz{-0.88cm}
\be
\z{$\zzzzzz{\White{\Theta}}^{\m\zz{\z{\yyy\tt\2L}}{}\m}\zzz{-0.475cm}
{\z{\zzzzz$\bar{\z{\uuuuu$\Theta$}}$}}_{\mm\zz{}{4\14}}$}
\1\defin\2  
\z{\x\zzzzz$\varepsilon$}  
\4=
\m-\1\w{\0\1y\m}{\2\0\z{\sf V}\1\2}\8\w{\2\partial\1\ln\0Z
(\0 \z{\sf\1V\m}\m\z{\zzzzz$/$}\1y\m=\m\z{\tt const}
\0=\0{x}^{\13}\1)\1}{\partial\2y}
\4=\4  
\w{\2\zzz{-0.18cm}\vx\xi\zzzz{-0.05cm}\vy\zzz{-0.18cm}\2}{\z{\sf V}}
\5
\z{\zzzzz$\langle$}\3
{\partial}_{\zzz{-0.01cm}\zz{}{\xi}}\3
{\z{\zzzzz$S$}}_{\mm\zz{}{\z{\yyy\sf G}}}\zzz{0.15mm}
\3
{\z{\zzzzz$\rangle$}}
\zzz{-0.245cm}\1\vx
\z{\yyy$a=\z{\yyy\sf const}$}\zzzz{0.6cm}
\vy
\zzz{-0.18cm}
\z{\7,\zzz{-0.28cm}\0}
\zzz{0.182cm}
\ee

\zzzz{-0.5cm}
\np
whereat \vw{\z{\sf V}} is the \z{\four\nnn\0volume} of one 
periodicity segment of the regarded aniso\nnn\1-\xyz tropic 
lattice and \vw{{\z{\zzzzz$S$}}_{\mm\zz{}{\z{\yyy\sf G}}}} 
is the employed lattice action for the pure gauge field 
sector.

In lieu of dimensional regularization of the 
continuum \vw{\zzzzzz 
{\Theta}^{\zzz{0.01cm}\zzz{0.05mm}\zz{\z{\yyy$\mu$}}{}}
_{\zzz{-0.01cm}\zzz{0.05mm}\zz{\z{\yyy$\mu$}}{}}
}, the 
\z{(\nn\0spatial\1)}\xyz

\yzy 

\zzzz{-1.04cm}\np  
lattice spacing \z{\1"\m$a$\2\n"\m} 
is the regulator in (6.5), used  in order to prepare the 
renormalization procedure, which involves the lattice 
counterpart\zyx

\zzzz{-0.99cm}\np 
\vww{\zzzzzz
{\beta}_{\m\zz{}{L}}
(\1g
\zzz{-0.21cm}\vx
\z{\yyy$\circ$}
\zzzz{-0.27cm}
\vy\zzz{-0.17cm}
)
\1=\mmm-\0a 
\4
\partial\1
\mnn 
g
\zzz{-0.21cm}\vx
\z{\yyy$\circ$}
\zzzz{-0.27cm}
\vy\zzz{-0.17cm}\0
\uuu{\5}{/}{-0.5} 
\partial\1a
} relative to the 
positive\2-\1sign Callan\1-\1Symanzik function\xyz

\zzzz{-0.95cm}\np
in continuum 
physics. The na\"ive average formulae (6.5) and (6.7) have a 
ther\1-\xyz

\zzzz{-0.92cm}\np
modynamic character and 
compare \vw{\zzzzzz 
\uuu{\5}{\langle}{0}\1\nnn
{\Theta}^{\zzz{0.01cm}\zzz{0.05mm}\zz{\z{\yyy$\mu$}}{}}
_{\zzz{-0.01cm}\zzz{0.05mm}\zz{\z{\yyy$\mu$}}{}}
\3\nnn\uuu{\5}{\rangle}{0}
} with a renormalization procedure\xyz

\zzzz{-0.93cm}\np
containing the lattice Lagrangian 
and \vw{\zzzzzz 
\langle\3
{\Theta}_{\mm\zz{}{4\14}}
\3\rangle
} with the corresponding lattice\xyz

\zzzz{-0.97cm}\np
Hamiltonian, being equal 
to \vw{\zzzzzz 
\langle\3
{\Theta}_{\mm\zz{}{4\14}}
\3\rangle
} in 
the sense pointed out above. The\xyz

\zzzz{-1cm}\np
choice (4.2) symbolically 
attaches standard Wilson form to the mentioned Lagrangian 
and Hamiltonian\1!

The same advantage is realized by the 
\vw{{\Theta}_{\mm\zz{}{\mu\1\nu}}} lattice model 
of the Karsch\xyz

\zzzz{-1.02cm}\np 
group,  
(4.3\2a) and (4.3\2b), 
but at the cost of renouncing a uniform construc\1-\xyz tion 
principle for its components. On the other hand, the 
\vw{{\Theta}_{\mm\zz{}{\mu\1\nu}}} lattice model\xyz

\zzzz{-0.99cm}\np  
of the Pisa group, 
(4.1), 
has a uniform construction principle but cannot generically 
offer a na\"ive Wilson access to the trace anomaly and the 
Hamil\nnn-\xyz tonian component \vw{\zzzzzz 
{\Theta}_{\mm\zz{}{4\14}}
}. The 
new ansatz presented here, 
(4.2), 
agrees upon\xyz

\zzzz{-0.99cm}\np
both advantages, but it numerically suffers from a 
more complicated algebra\1-\xyz 

\zzzz{-0.99cm}\np
ic construction principle. There 
are no conservation laws for 
\vw{{\Theta}_{\mm\zz{}{\mu\1\nu}}} 
on the lat\1-\xyz 

\zzzz{-0.86cm}\np
tice\nnn\1--\nnn\1except for (4.1), 
where the 
\z{\n\0SU(N)\n} version 
of \vw{\zzzzzz
{\z{\x$\Theta$}}
^{\zz{\z{\yyy\sf c\nnn\0a\nnn\0r\nnn\0a\1\bf HYB\mn\0RID}}{}}
_{\mm\n\zz{}{\mu\1\nnnn\nu}}
\1
} is 
conserved\xyz

\zzzz{-0.98cm}\np
perturbatively 
in \z{\11\1-\1loop} order.

\SNN{Acknowledgements}
\label{Sec_Acknowledgements}

The author wishes to express his gratitude to H. J. Rothe for 
several fruit\1-\xyz ful discussions regarding applied lattice 
physics. Moreover it is a pleasure to thank I. O. Stamatescu, 
I. Bender, and J. Pol\'{o}nyi for further valuable suggestions.

%

%

\end{document}